\documentclass{aa} 

\usepackage{graphicx}
\usepackage{txfonts}
                                
\usepackage{color}
\newcommand{\Ms}{{\ensuremath{M_{\odot} }}}

\begin{document}

   \title{Radio emission from little red dots may reveal their true nature}

   \subtitle{}

    \author{Muhammad A. Latif\inst{1}, Ammara Aftab\inst{1}, Daniel J. Whalen\inst{2}, Mar Mezcua \inst{3,4}
          }
    \institute{Physics Department, College of Science, 
            United Arab Emirates University, PO Box 15551, Al-Ain, UAE \\
              \email{latifne@gmail.com}
         \and
             Institute of Cosmology and Gravitation, Portsmouth University, Dennis Sciama Building, Portsmouth PO1 3FX
         \and
             Institute of Space Sciences (ICE, CSIC), Campus UAB, Carrer de Magrans, 08193 Barcelona, Spain
         \and
              Institut d’Estudis Espacials de Catalunya (IEEC), Edifici RDIT, Campus UPC, 08860 Castelldefels (Barcelona), Spain
             }

   \date{}

 
  \abstract
   
   {The unprecedented sensitivity of the \textit{James Webb Space Telescope} (\textit{JWST}) has revolutionized our understanding of the early universe. Among the most intriguing \textit{JWST} discoveries are red, very compact objects showing broad line emission features nicknamed as little red dots (LRDs). The discovery of LRDs has triggered great interest about their origin as either extremely starbursting galaxies or highly-obscured active galactic nuclei (AGN). Their exact nature still remains unknown.}
  {The goal of this work is to estimate the radio emission from LRDs and predict which radio surveys would detect them. To achieve these objectives, we employ the fundamental plane of black hole (BH) accretion to estimate radio emission from AGN and the stellar radio fluxes from their host galaxies. We assume a range of BH mass, X-ray luminosity ($\rm L_{X}$) and star formation rate (SFR) to bracket the likely properties of LRDs.}
  {Our findings suggest that BH radio fluxes from LRDs are 10-100 times higher than the stellar fluxes from their host galaxies, depending on BH mass, $\rm L_X$ and SFR. The detection of a $\sim$ 500 nJy signal above 2 GHz at $z \geq$ 5 or a $\sim$ 2000 nJy flux at $z =$ 3-4 would be a smoking gun for the presence of AGN provided that SFRs in the host galaxies are $\rm < 30~ \Ms ~yr^{-1}$.}
  {We find that LRDs are most likely radio quiet AGN otherwise would have been already detected in the current radio surveys. Our findings suggest that LRDs can be detected with the upcoming radio observatories such as ngVLA and SKA with integration times of 10-100 hrs, respectively.}
  
   {}

\titlerunning{Radio Emission from LRDs}
\authorrunning{Latif et al.}
   \maketitle
%

\section{Introduction}

 Recent \textit{JWST} photometric and spectroscopic surveys have uncovered a large sample of faint AGN candidates at $z =$ 4-10 that have led to a seismic shift in our understanding of the AGN population \citep{Bod23,Kok23,Furt23,Lamb23,Lars23,Gould23,Maio23a,Mat24,Koko24,G24}. Among the most intriguing discoveries are very compact, reddish objects exhibiting broad line emission features in their spectra, the LRDs \citep{Mat24,G24,Koc24,Mai24}. Their spectral energy distributions (SEDs) show v-shape features that can be explained by extremely starbursting galaxies, highly-obscured AGN, or a combination of the two. If LRDs are due to rapid star formation (SF), their inferred stellar masses would be $\rm 10^{10}-10^{11}~\Ms$, unusually large for galaxies at such redshifts \citep{Perez24}. Although supernova (SN)- or SF-generated outflows may explain observational features of LRDs such as broad line emission, narrow unresolved forbidden lines such as [O III] cannot be explained by just stars \citep{Mai24}. Alternatively, they could be heterogeneous galaxies with contributions both from stars and AGN.

Spectroscopic studies suggest that LRDs are dust-reddened AGN with $L_{\rm bol}=10^{43-46}$ $\rm erg~s^{-1}$ and inferred BH masses of $\rm 10^6-10^8$  \Ms\ \citep{G24,Mat24}. Moreover, their cosmic abundance is two orders ($\rm \sim 10^{-5} Mpc^{-3} mag^{-1}$) of magnitude higher than quasars observed at such redshifts \citep{Mat24,Koko24,G24}. LRDs provide key insights about the AGN population, co-evolution of AGN with their host galaxies, BH seeding and growth mechanisms as well as AGN contribution to cosmic reionization. So far, near infrared (NIR) observations by \textit{JWST} remain inconclusive as both SF and AGN can produce similar spectral features. X-ray emission, which could be a robust indicator of the presence of AGN, has only been detected for a few LRDs to date \citep{Mai24,Koc24} and mostly remains undetected \citep{Ana24,Yue24}. Nevertheless, these studies provide useful constraints on AGN activity such as their BH masses and accretion rates. Multi-wavelength observations may unambiguously reveal the signatures of AGN activity. However, while both NIR and X-rays are obscured by dust in the host galaxies, radio is not. \cite{Ak24} report non-detections of LRDs in radio in current data from Meerkat at 1.4 GHz and VLA at 3 GHz, potentially ruling out that the average LRDs are radio-loud AGN. Deeper future observations are required to determine their properties, and their true nature. 

Motivated by the prospects of detection of $\gtrsim 10^6$ \Ms\ BHs by upcoming radio surveys by the Next-Generation Very Large Array (ngVLA) and the Square Kilometer Array (SKA) up to $z \lesssim 13 - 14$ \citep{W23,L24b,L24} we propose here following up LRDs in radio. In Section 2 we lay out our method for estimating radio fluxes from AGN and host galaxies. We present our flux estimates in Section 3 and discuss our conclusions in Section 4.

\section{Numerical Method} \label{sec:method}

\subsection{Radio emission from BHs in LRDs}

We estimate the radio flux  at 0.1 - 10 GHz from BHs in LRDs with the fundamental plane (FP) of BH accretion. The FP is an empirical relation  between the nuclear radio luminosity ($L_\mathrm{R}$) at 5 GHz, the BH mass ($M_\mathrm{BH}$) and the X-ray luminosity ($L_\mathrm{X}$) at 2 - 10 keV \citep{merl03}. They are valid over five orders of magnitude in BH mass down to $M_\mathrm{BH} \sim 10^5$ \Ms\ \citep{gul14}. To compute the expected radio flux from the BH in the observer frame, we calculate $L_\mathrm{R}$ from $L_\mathrm{X}$ with the FP:

\begin{equation}
\mathrm{log} \, L_\mathrm{R} (\mathrm{erg \, s^{-1}})= A \, \mathrm{log} \, L_\mathrm{X} (\mathrm{erg \, s^{-1}}) + B \, \mathrm{log} \, M_\mathrm{BH} (\mathrm{M}_{\odot})+ C,
\end{equation}
where A, B, and C are taken from \citet{merl03}, \citet{kord06}, \citet{gul09}, \citet{plot12}, and \citet{bonchi13} and are listed in Table 2 of \cite{L24b}. 

\cite{B22} found that the FP depends on the radio loudness of the source. To take this into account we use the relations for radio-quiet and radio-loud sources from \cite{B22}. For radio-loud cases, we use $A=1.12 \pm 0.06$, $B=0.20 \pm 0.07$, and $C=- (5.64 \pm 2.99)$ and for the radio-quiet cases, $A=0.48 \pm 0.06$, $B=0.50 \pm 0.08$, and $C=15.26 \pm 2.66$. X-ray observations of LRDs provide upper limits on their X-ray luminosities at 2-10 KeV \citep{Ana24,Yue24}. We take three values of $L_\mathrm{X}=$ $\rm 10^{43}$ erg s$^{-1}$, $\rm 10^{43.5}$ erg s$^{-1}$ and $\rm 10^{44}$ erg s$^{-1}$ from table 1 of \cite{Ana24}. Similarly, three fiducial values of $M_\mathrm{BH} =$ $\rm 10^{6} ~\Ms$, $\rm 10^{6.5} ~\Ms$, and $\rm 10^{7} ~\Ms$ are taken based on the X-ray observations of LRDs mentioned above.

To cover the range of LRD redshifts we consider $z=3-7$.  Because of cosmological redshifting, the radio flux does not originate from 5 GHz in the observer frame today. Radio fluxes from the FP are redshifted into ngVLA and SKA bands today using the relation $L_\mathrm{R} =$ $\nu L_{\nu}$ in the source frame, assuming that $L_{\nu} \propto \nu^{-\alpha}$. As explained in Sect. 4, we use a spectral index $\alpha$ = 0.3 and compute the spectral flux at frequency $\nu$ in the observer frame from the spectral luminosity at $\nu'$ in the source frame from
\begin{equation}
F_\nu = \frac{L_{\nu'}(1 + z)}{4 \pi {d_\mathrm L}^2},
\label{eq:flx}
\end{equation}
where $d_\mathrm L$ is the luminosity distance and $\nu' = (1+z) \nu$.  We employ cosmological parameters from \textit{Planck}:  $\Omega_{\mathrm M} = 0.308$, $\Omega_\Lambda = 0.691$, $\Omega_{\mathrm b}h^2 = 0.0223$, $\sigma_8 =$ 0.816, $h = $ 0.677 and $n =$ 0.968 \citep{planck2}.

\subsection{Radio Emission from host galaxies}

Massive stars in host galaxies create H II regions that produce continuum radio flux via thermal bremsstrahlung emission. Also the synchrotron emission from young supernova remnants (SNRs) in host galaxies produces radio flux that contributes to the total radio emission from AGN, particularly at lower frequencies. To estimate the radio emission from host galaxies, we use the empirical relation given in \cite{D17} that also takes into account the redshift dependence which we briefly summarize here (see \cite{D17} for further details). We estimate the radio luminosity in the following way:
\begin{equation}
L_R [W/Hz] = \frac{SFR[\Ms/yr]}{f_{IMF}10^{-24}10^{q_{TIR}(z,\alpha)}}
\label{eq:lr}
\end{equation}
where
\begin{equation}
q_{TIR}(z,\alpha) = (2.88 \pm 0.03)(1 + z)^{-0.19 \pm 0.01}
\label{eq:tir}
\end{equation}
and $f_{IMF}$ is the factor for the assumed initial mass function (IMF) which we take $f_{IMF}=1.7$ for the Salpeter IMF. $\alpha$ is the assumed spectral index of the star-forming galaxy which we take to be - 0.7.
We estimate the stellar flux using the following relation:
\begin{equation}
f_{\nu_{obs}} = \frac{L_R (1+z)^{\alpha+1}}{4 \pi {d_\mathrm L}^2} \left( \frac{1.4}{\nu_{obs}}\right)^{\alpha} 
\label{eq:lr}
\end{equation}
We take two values of the observing frequency ($\nu_{obs}$) 3 GHz and 8 GHz and $\alpha$= - 0.7. The above relation has been tested by \cite{D17} for 3 GHz but we assume it to be valid for 8 GHz, this assumption does not impact our results. SFRs in the host galaxies of LRDs are not directly measured but are typically estimated with stellar synthesis codes and remain largely uncertain because of degeneracies in the models \citep{G24}. To bracket the range of SFRs inferred for LRDs, we consider 1, 10, and 30 $\Ms$ yr$^{-1}$ at each redshift.

\section{Results} \label{sec:res}

\begin{figure*} 
\begin{center}
\includegraphics[scale=0.35]{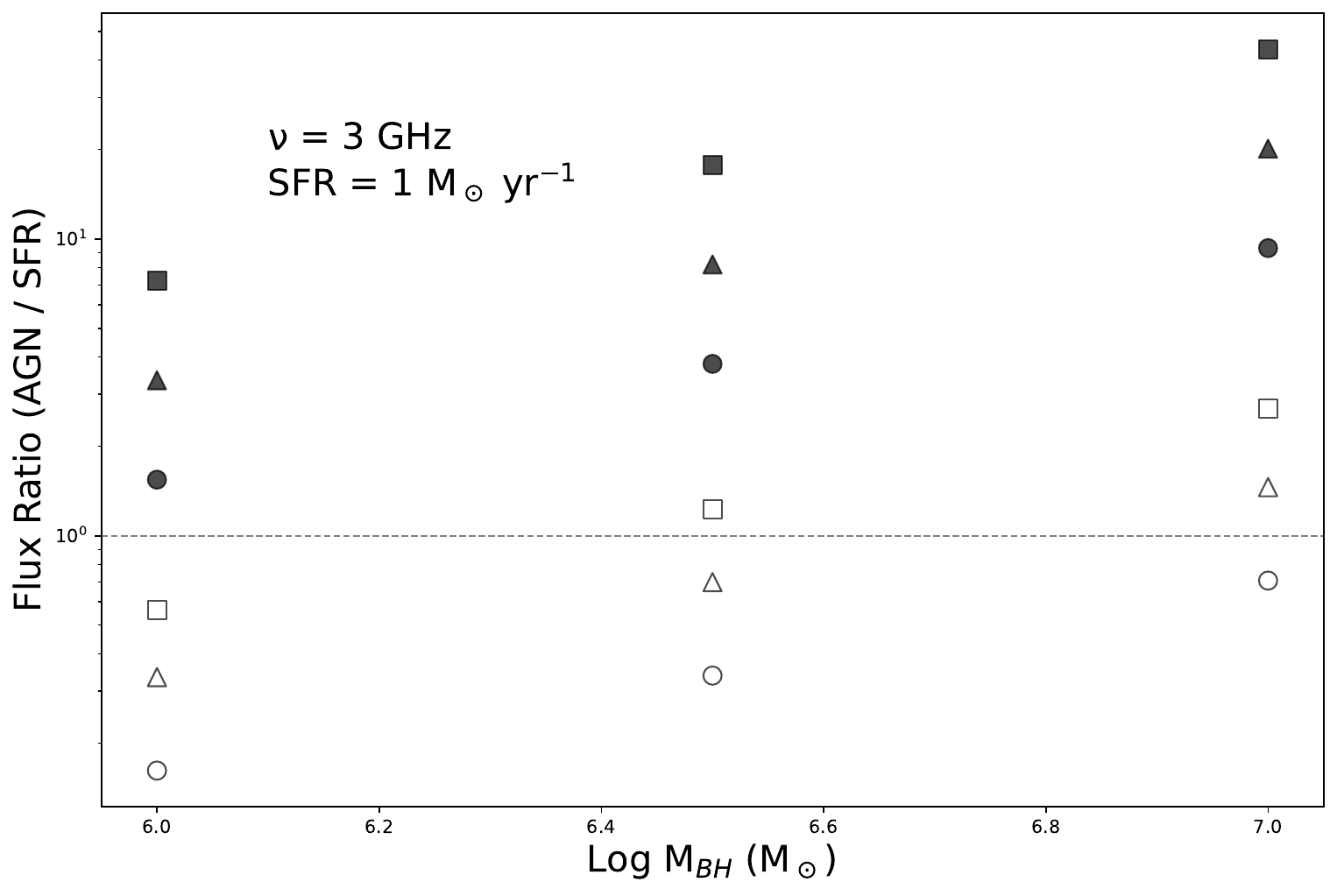} 
\includegraphics[scale=0.35]{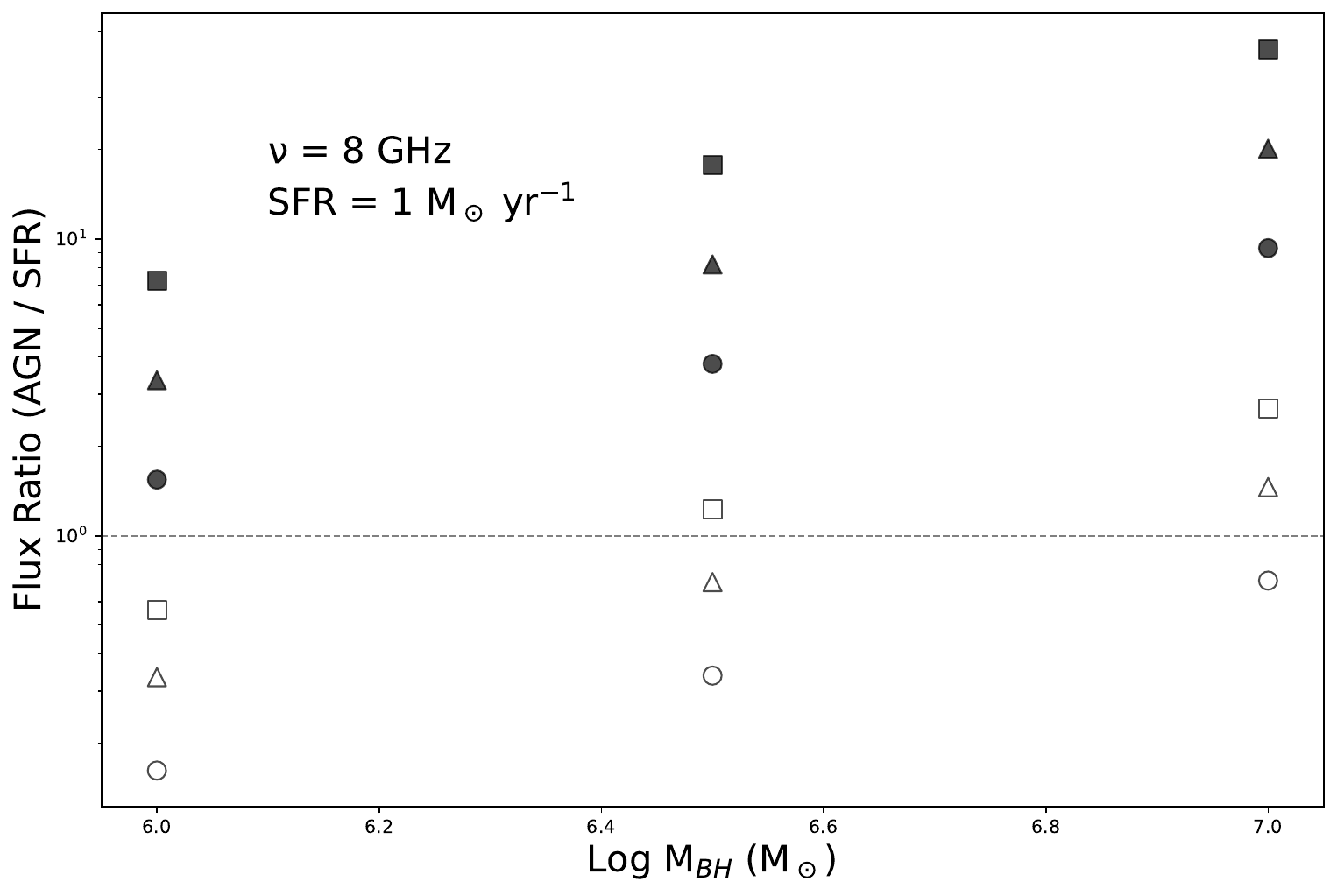} 
\includegraphics[scale=0.35]{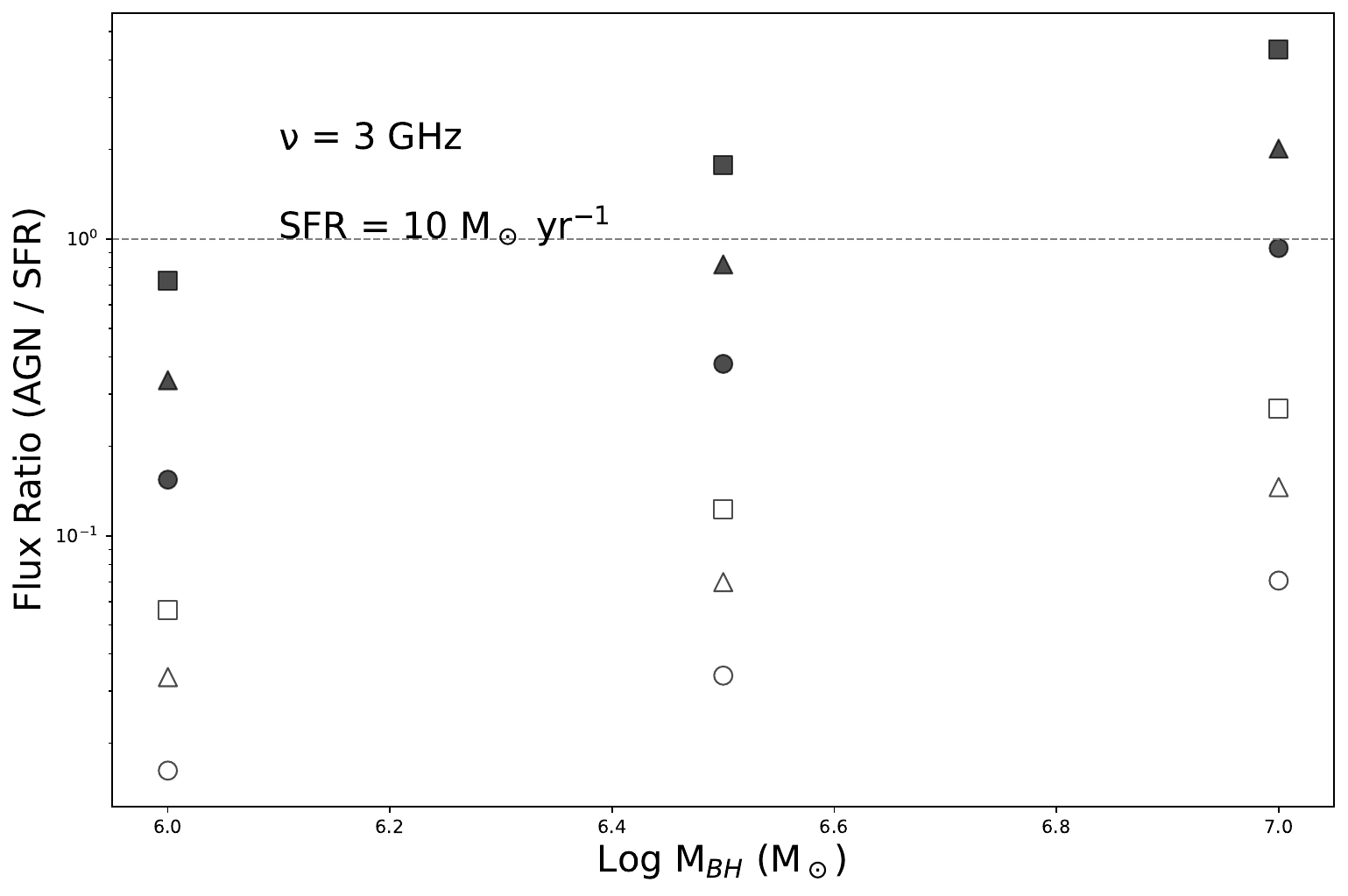} 
\includegraphics[scale=0.35]{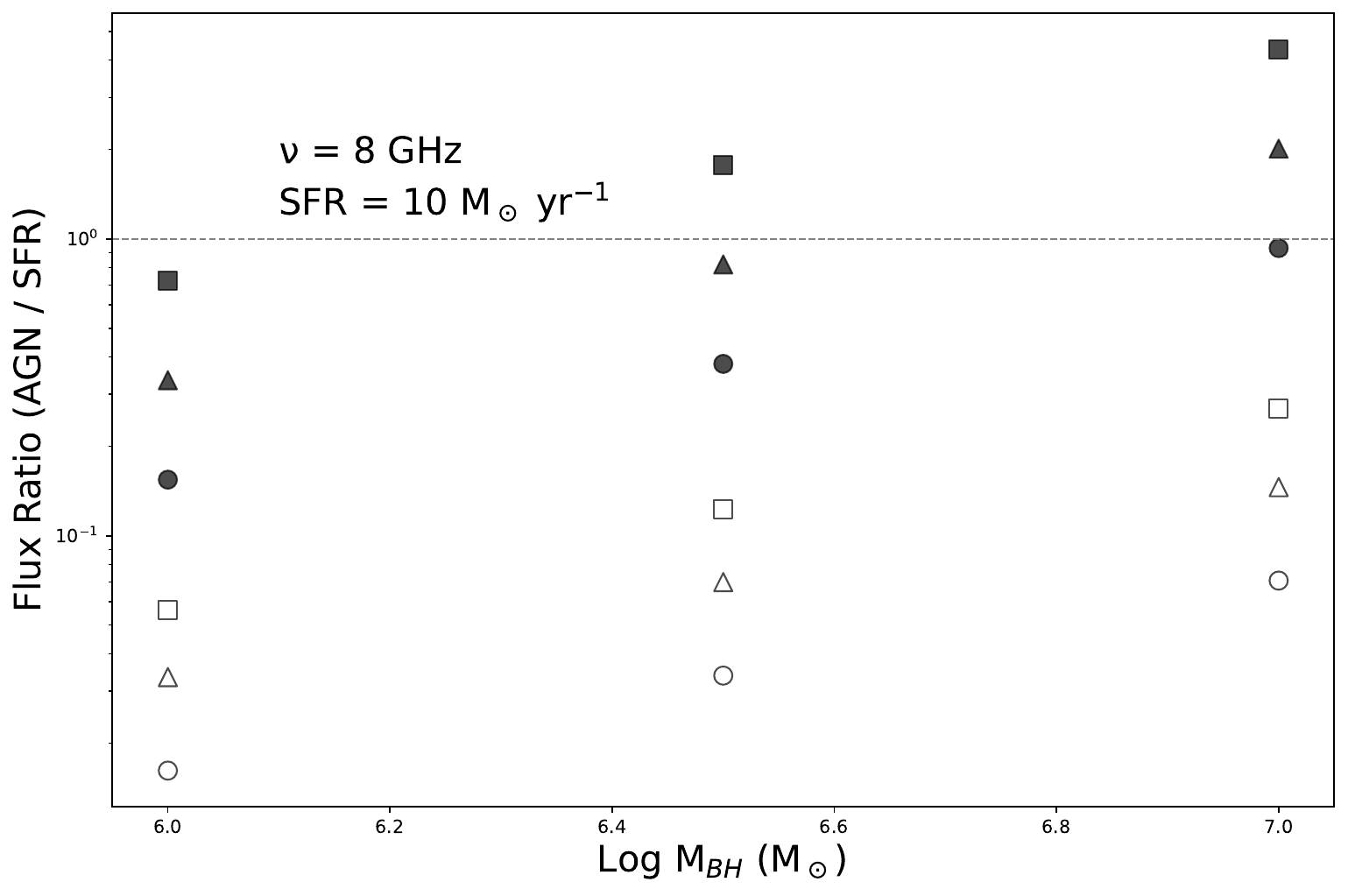} 
\includegraphics[scale=0.35]{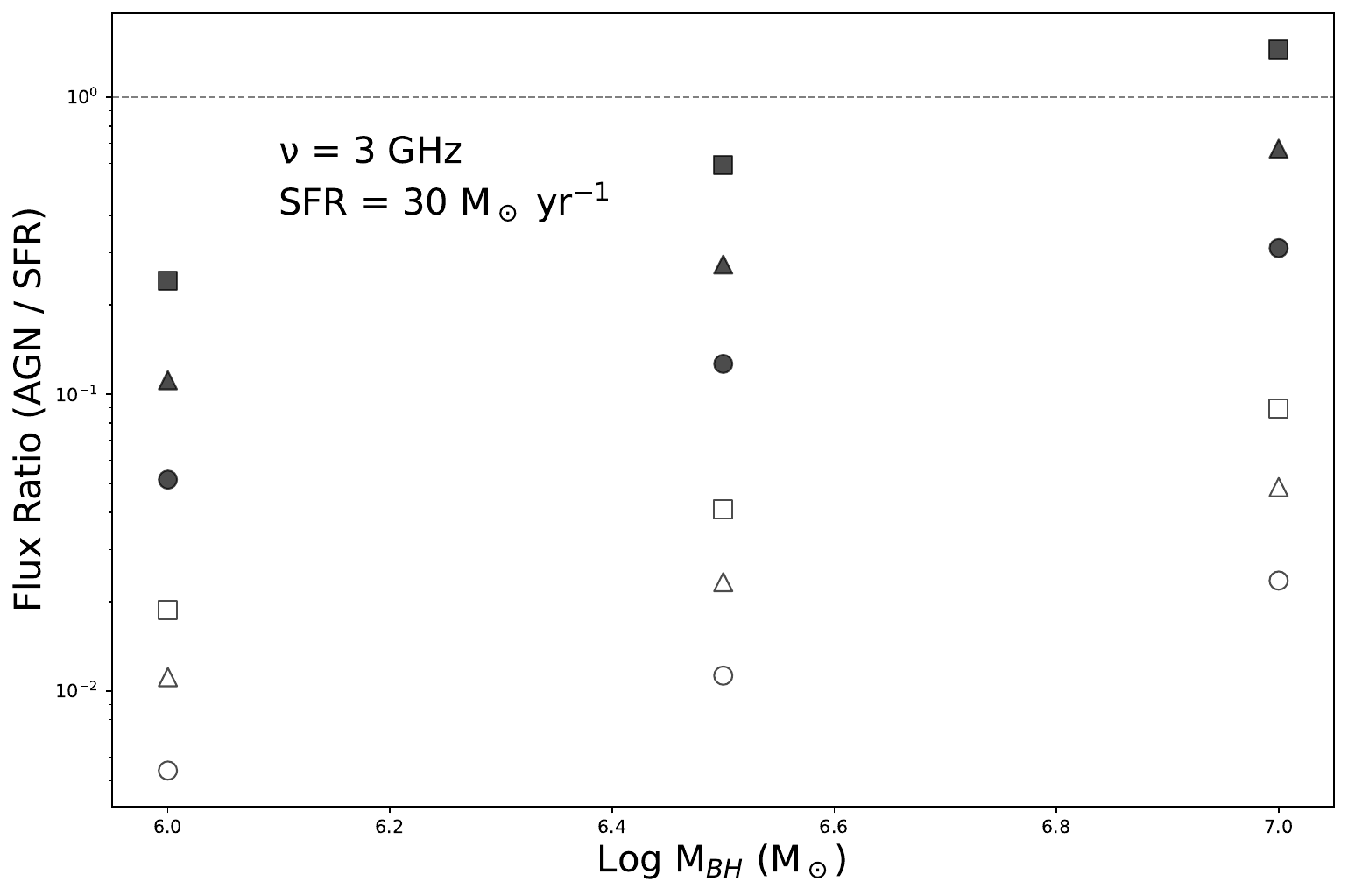} 
\includegraphics[scale=0.35]{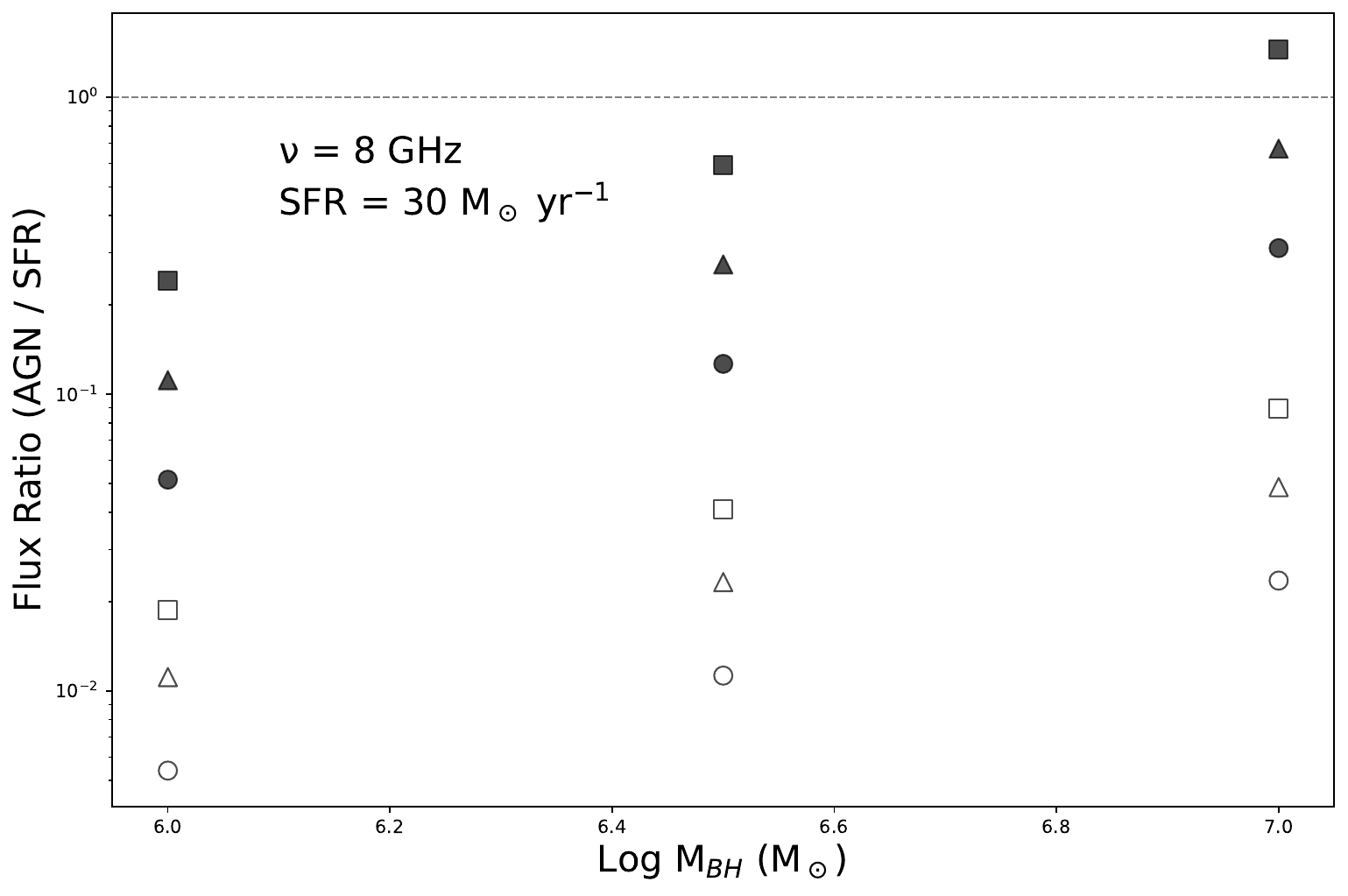} 
\end{center}
\vspace{-0.1cm}
\caption{AGN to SFR flux ratios are shown vs BH mass at $z =$ 7. The empty and solid symbols are for the minimum and the maximum fluxes from FP, respectively. Circles, triangles and squares are for log($L_{\rm X})$ = 43 erg ~s$^{-1}$, log($L_{\rm X})$ = 43.5 erg ~s$^{-1}$ and log($L_{\rm X})$ = 44 erg ~s$^{-1}$, respectively. Horizontal dashed lines are for AGN/SFR flux ratio = 1.}
\label{f1}
\end{figure*}

The AGN to SFR flux ratios vs BH masses are shown for 3 GHz and 8 GHz at $z =$ 7 in Fig. \ref{f1}. These two frequencies are best for distinguishing AGN fluxes from those of their host galaxies \citep{W23,L24b,L24}. Also, they allow us to compare our results with the VLA Cosmic Evolution Survey (COSMOS). We here consider two scenarios: with and without radio loudness. We label the first case as the standard scenario and the second as the radio-loudness scenario.

\subsection{Standard scenario}
We find that for SFR $=$ 1 \Ms\ yr$^{-1}$, the flux ratios vary from 0.1 - 40, the maximum AGN fluxes are dominant over the host galaxies for the entire range of $M_{BH}$ and $L_{\rm X}$ but the minimum AGN fluxes only exceed the host galaxies for $\rm M_{BH} > 10^{6.5}~ \Ms$ and  $L_{\rm X} \geq 10^{43.5} \rm ~erg~ s^{-1}$. For SFR $=$ 10 \Ms\ yr$^{-1}$ the flux ratios vary from 0.02 - 4, the minimum AGN fluxes are subdominant compared to the emission from host galaxies for all cases and the maximum AGN fluxes only exceed stellar fluxes for $\rm M_{BH} > 10^{6.5}~ \Ms$ and  $L_{\rm X} \geq 10^{43.5} \rm ~erg~ s^{-1}$. In the case of SFR $=$ $30~\Ms$ yr$^{-1}$ flux ratios drop to 0.002 - 2, stellar fluxes from host galaxies are higher than both minimum and maximum fluxes only with the exception of $\rm M_{BH} = 10^7$ \Ms with $L_{\rm X} \geq 10^{44} \rm ~erg~ s^{-1}$. We show flux ratios for $z=7$ in Fig. \ref{f1}, they actually increase at lower redshifts. Also, the flux ratios at 8 GHz do not significantly differ from those at 3 GHz.

\begin{figure*} 
\begin{center}
\includegraphics[scale=0.3]{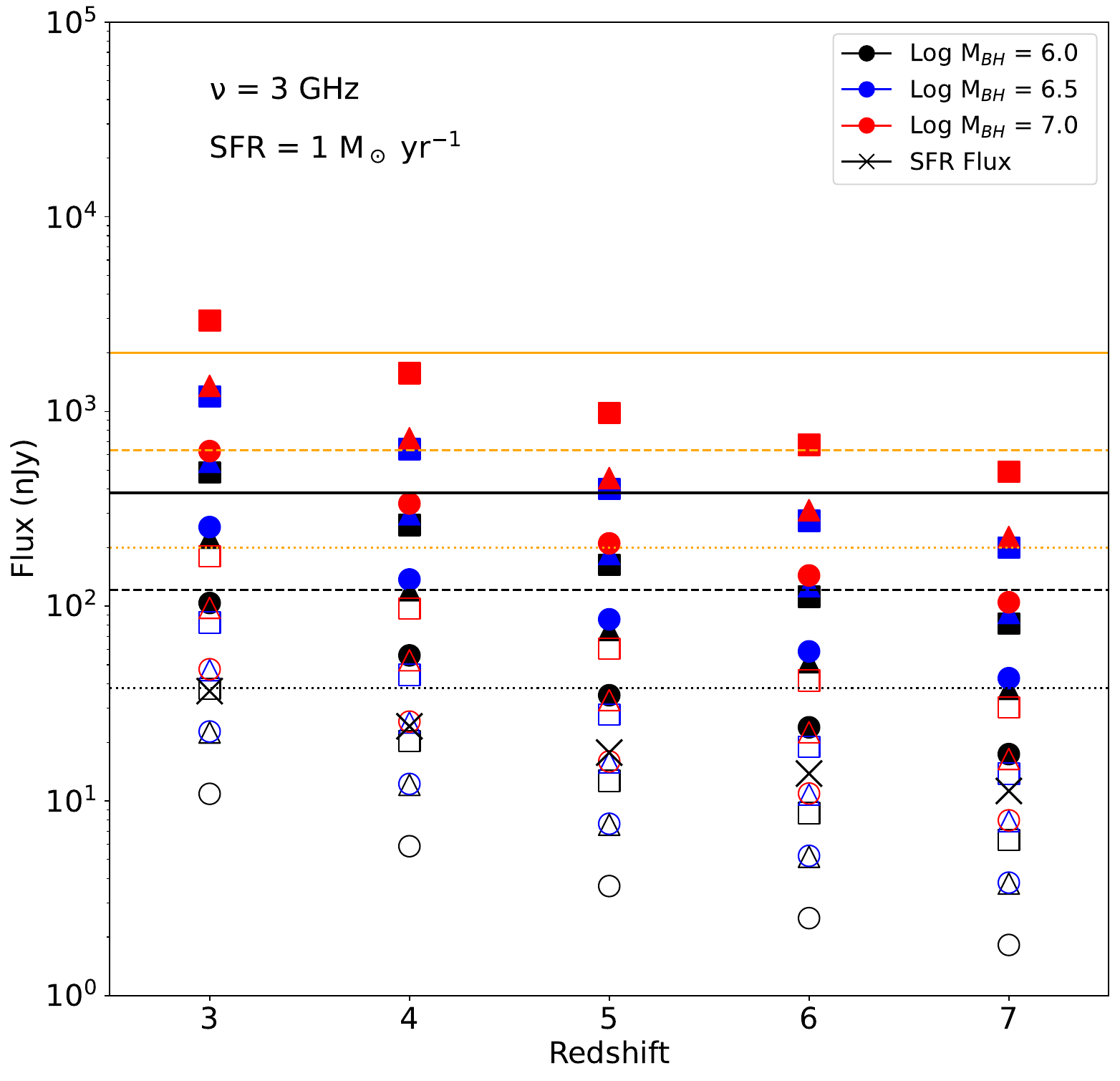} 
\includegraphics[scale=0.3]{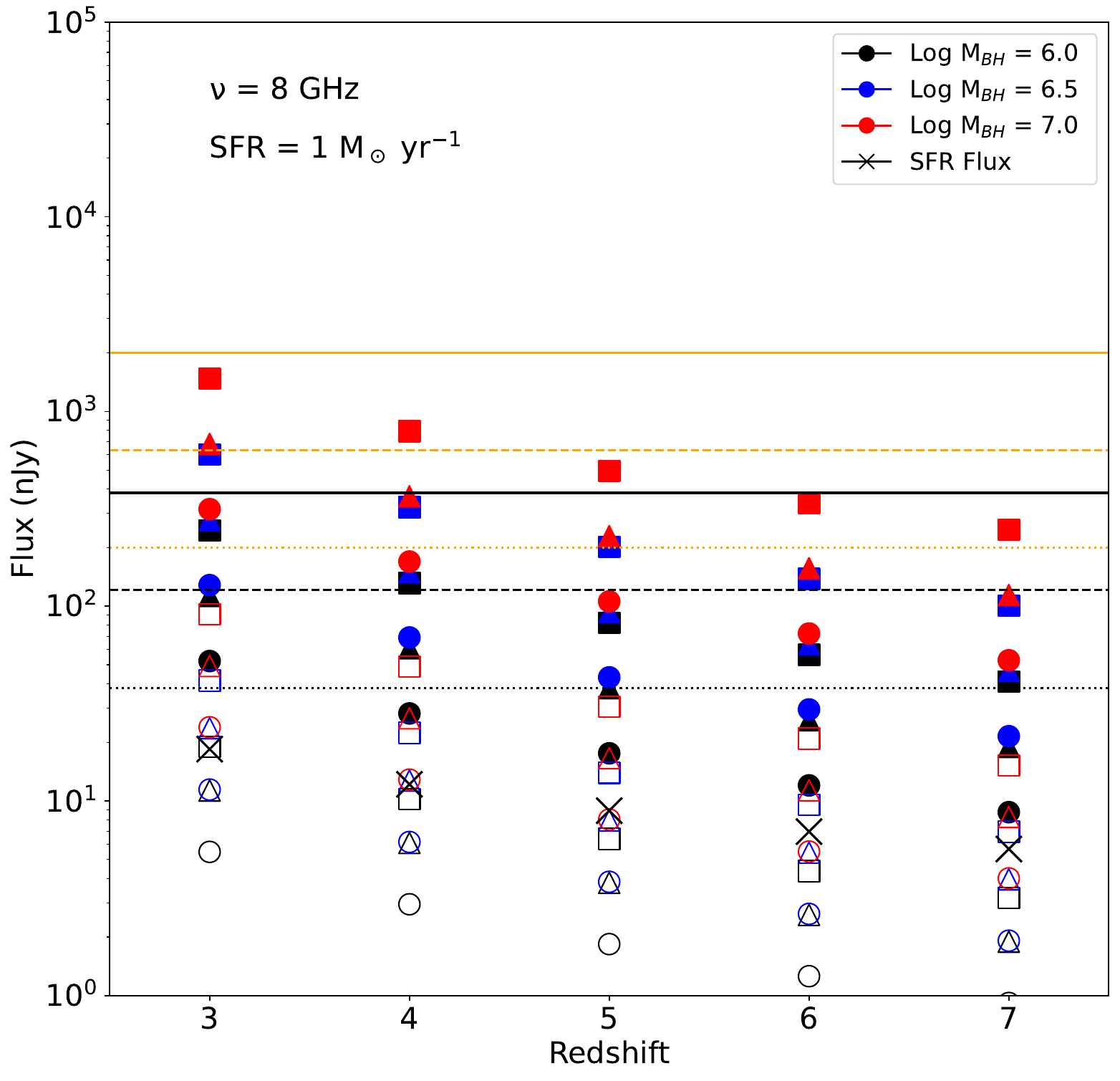} 
\includegraphics[scale=0.3]{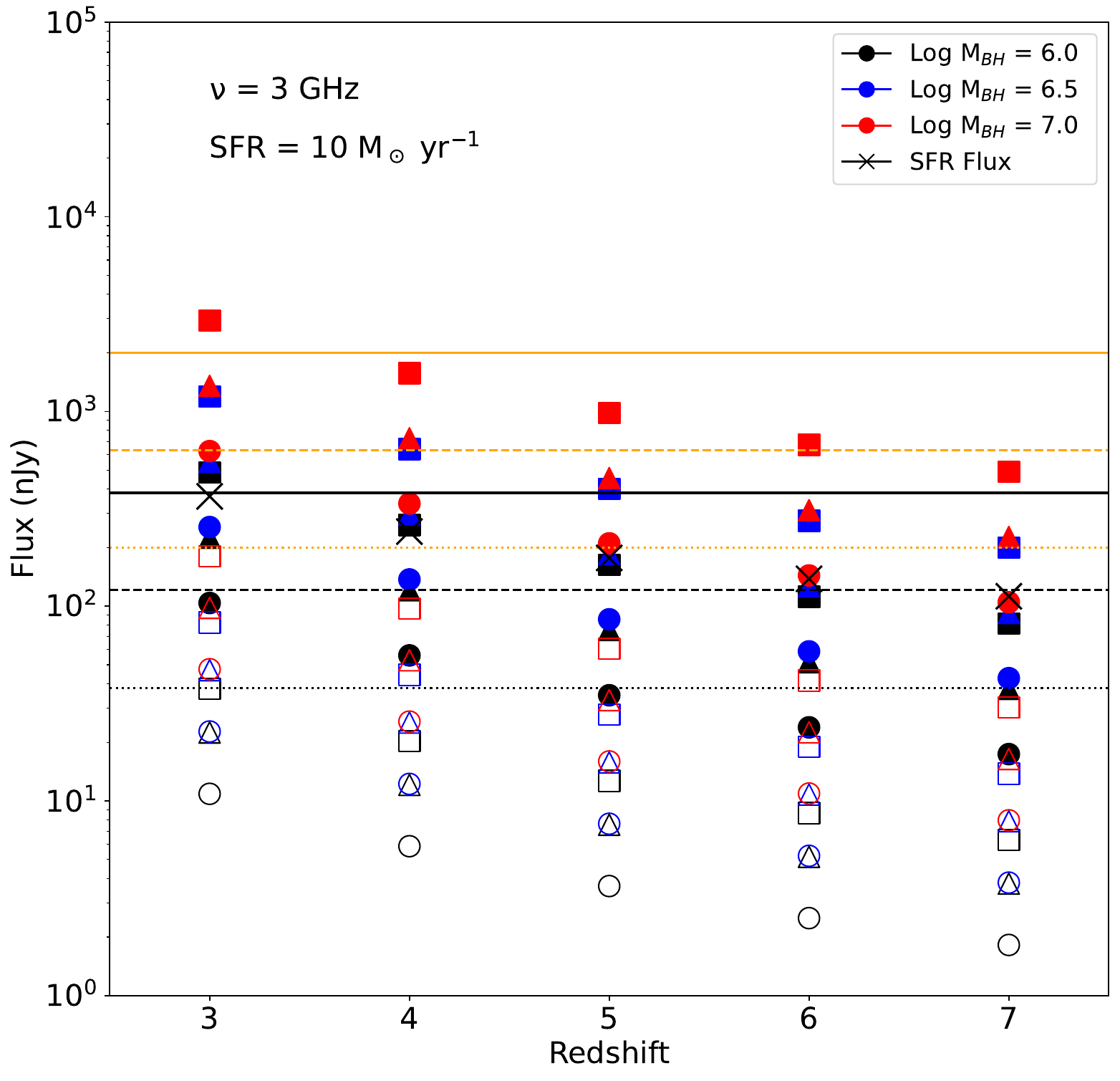} 
\includegraphics[scale=0.3]{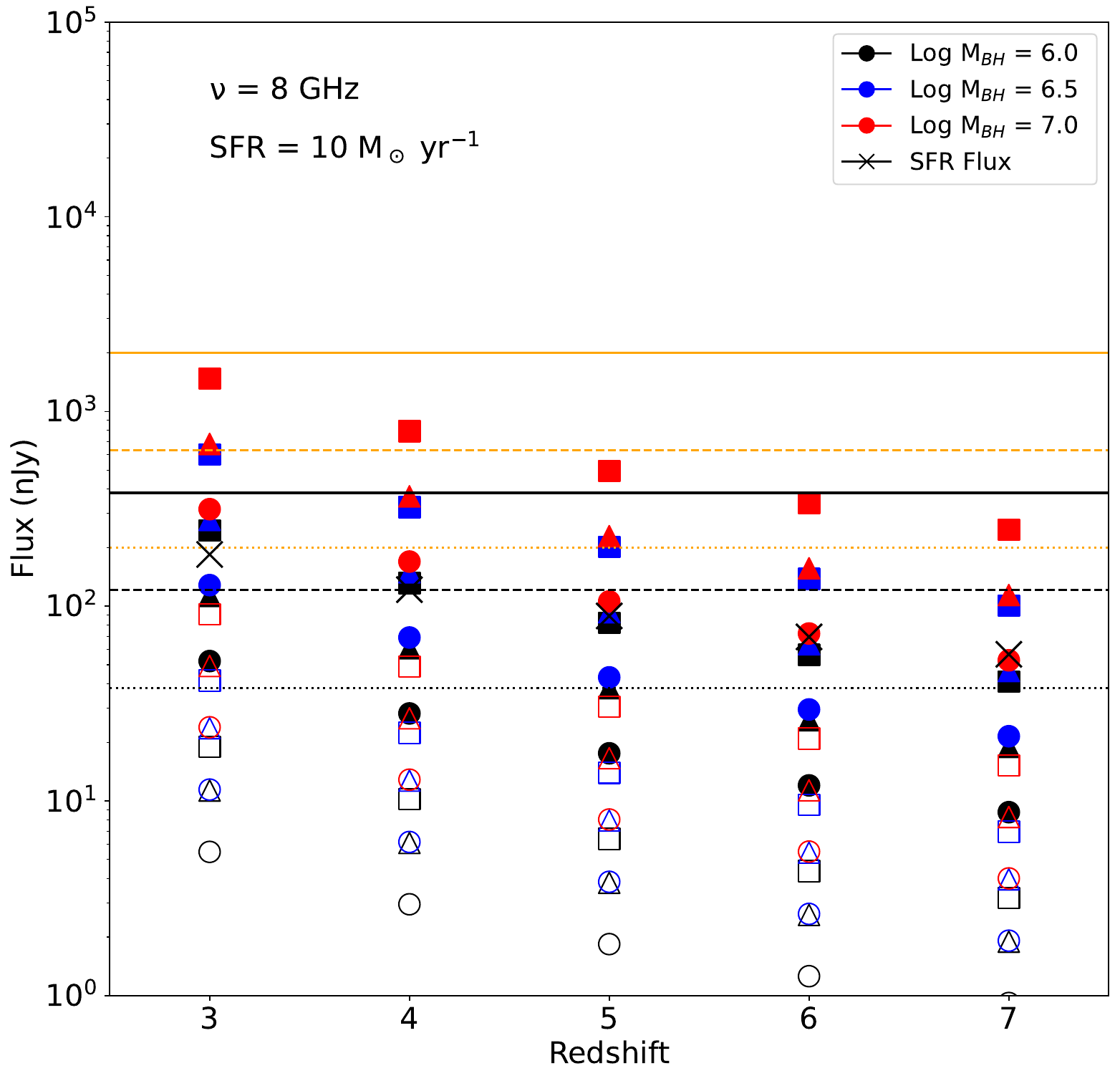} 
\includegraphics[scale=0.3]{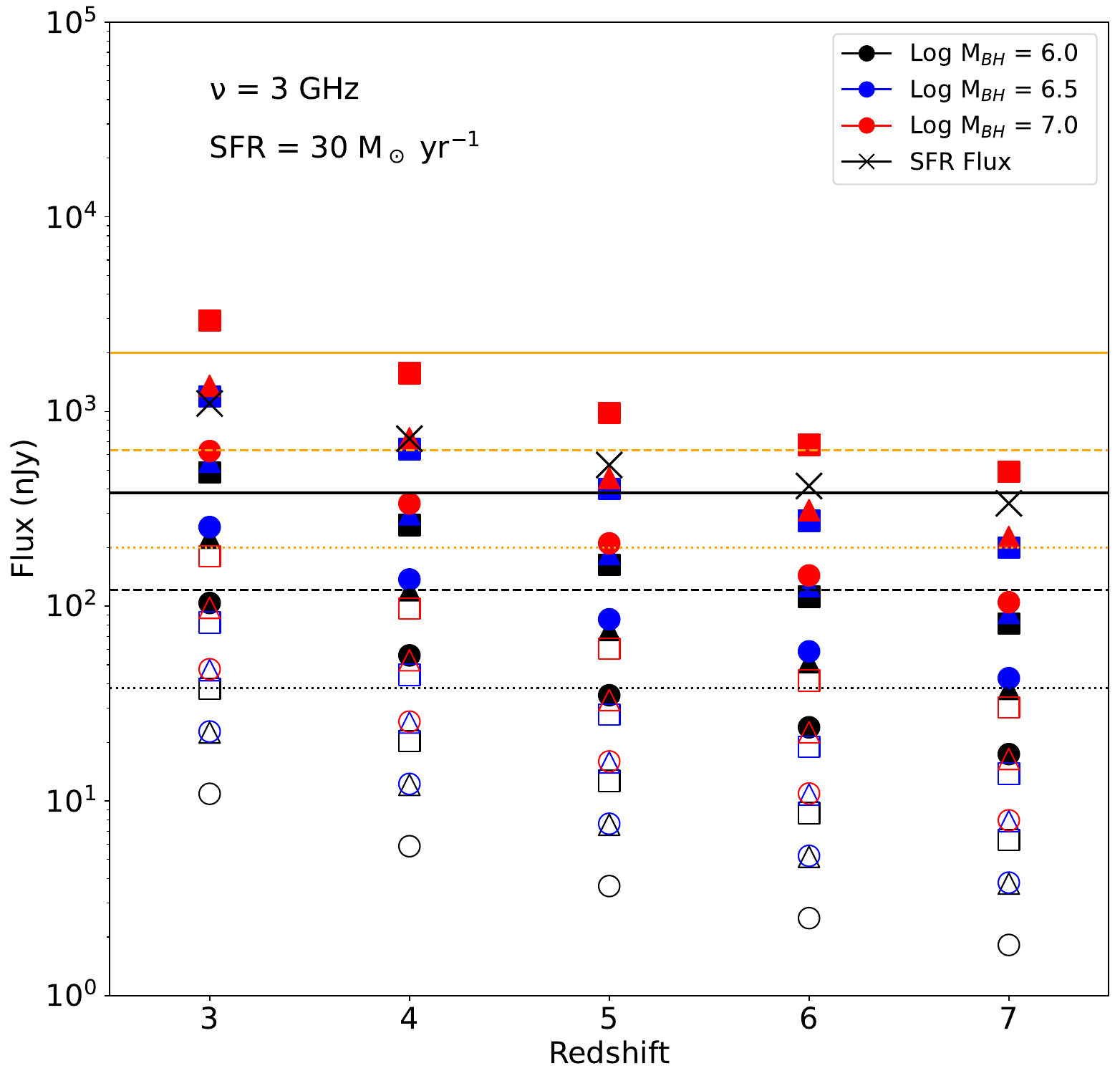} 
\includegraphics[scale=0.3]{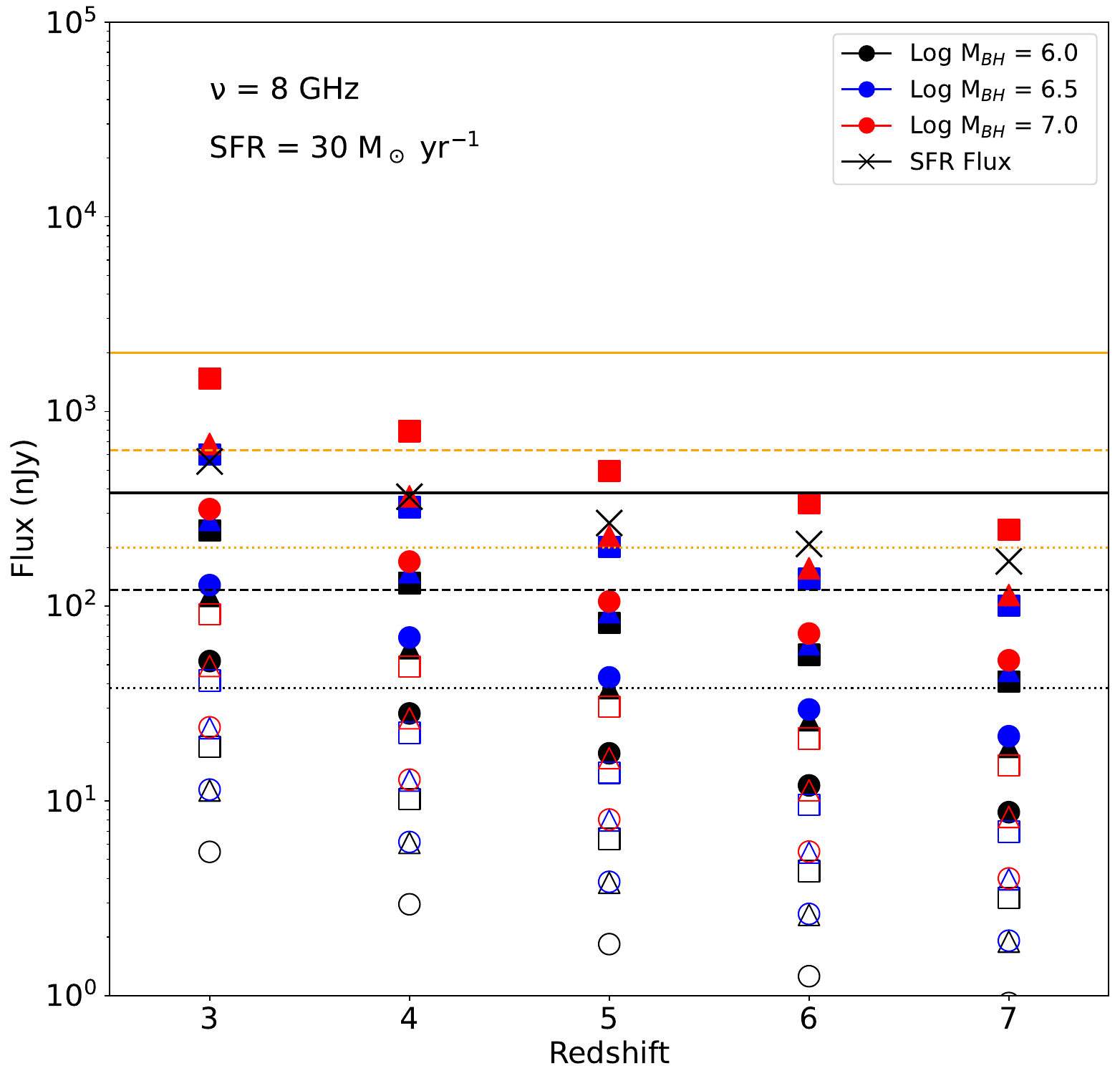} 
\end{center}
\vspace{-0.1cm}
\caption{Fluxes vs redshift. Empty shapes are for the minimum fluxes and filled shapes for the maximum fluxes from FP. Colors indicate BH mass and black crosses show SFR fluxes.  Circles, triangles and squares are for log($L_{\rm X})$ = 43 erg ~s$^{-1}$, log($L_{\rm X})$ = 43.5 erg ~s$^{-1}$ and log($L_{\rm X})$ = 44 erg ~s$^{-1}$, respectively. The orange and black lines are sensitivity limits for SKA and ngVLA, respectively, for integration times of 1 hr (solid), 10 hr (dashed) and 100 hr (dotted).}
\label{f2}
\end{figure*}

The fluxes for AGN and stars in their host galaxies are shown at $3 < z < 7$ in Fig. \ref{f2}. For brevity, only the minimum and maximum AGN fluxes (associated with the error bars of the FP) for $\alpha =$ 0.3 are shown. The six FP fluxes vary by a factor of about 20. The AGN radio fluxes range from 1 - 3000 nJy and decrease by a factor of a few from $z =$ 3 - 7. On the other hand, the stellar fluxes from the host galaxies vary from 10 - 40 nJy, 100 - 400 nJy and 400 - 1000 nJy for SFRs $=$ 1, 10 and 30 $\Ms$  yr$^{-1}$, respectively. The AGN radio fluxes are highest at $z =$ 3 for $M_{\rm BH} = 10^7$ \Ms and $L_{\rm X} = 10^{44} \rm ~erg~ s^{-1}$ and lowest at $z =$ 7 for $M_{\rm BH} = 10^{6}$ \Ms and $L_{\rm X} = 10^{43} \rm ~erg ~s^{-1}$. Generally, the maximum AGN fluxes are higher than those of the host galaxies for SFR $=$ 1~\Ms\ yr$^{-1}$ for all cases while the minimum AGN fluxes only exceed host galaxies at $M_{\rm BH} = 10^{7}$ \Ms\ and $L_{\rm X} = 10^{44} \rm ~erg~ s^{-1}$. 
For SFR $=$ 10~\Ms\ yr$^{-1}$, the maximum AGN fluxes only exceed the stellar fluxes from the host galaxies for  $M_{\rm BH} = 10^{7}$~\Ms\ at z<5. The minimum AGN fluxes are lower than the stellar fluxes from the host galaxies for all cases. For SFR $=$ 30~\Ms\ yr$^{-1}$, only the maximum AGN flux from $M_{\rm BH} = 10^{7}$~\Ms\ with $L_{\rm X} = 10^{44} \rm ~erg ~s^{-1}$ is higher than the stellar fluxes while in all the remaining cases the stellar fluxes from the host galaxies outshine their AGN. These trends are found at all redshifts.

\subsection{Radio-loudness scenario}
\begin{figure*} 
\begin{center}
\includegraphics[scale=0.35]{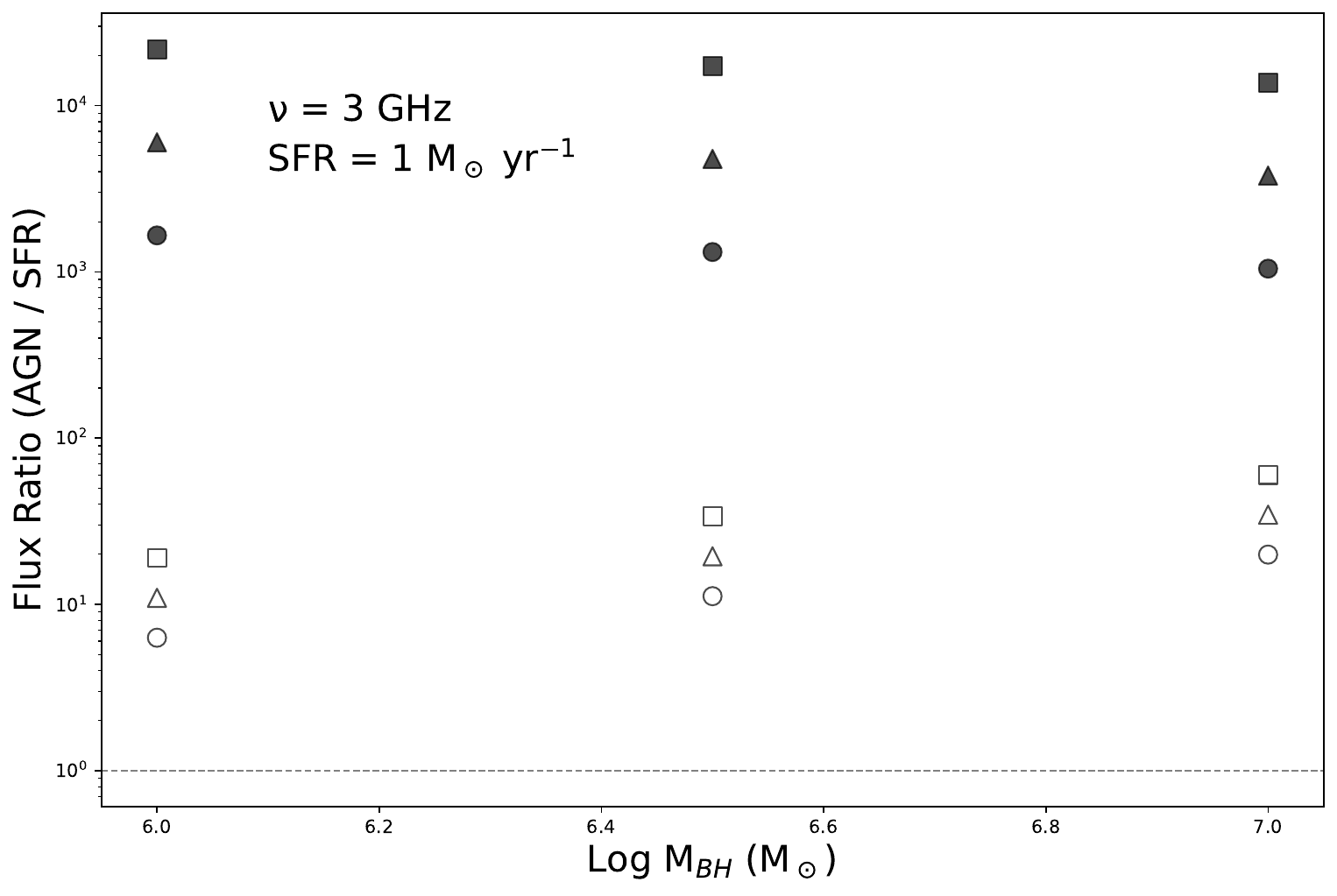} 
\includegraphics[scale=0.35]{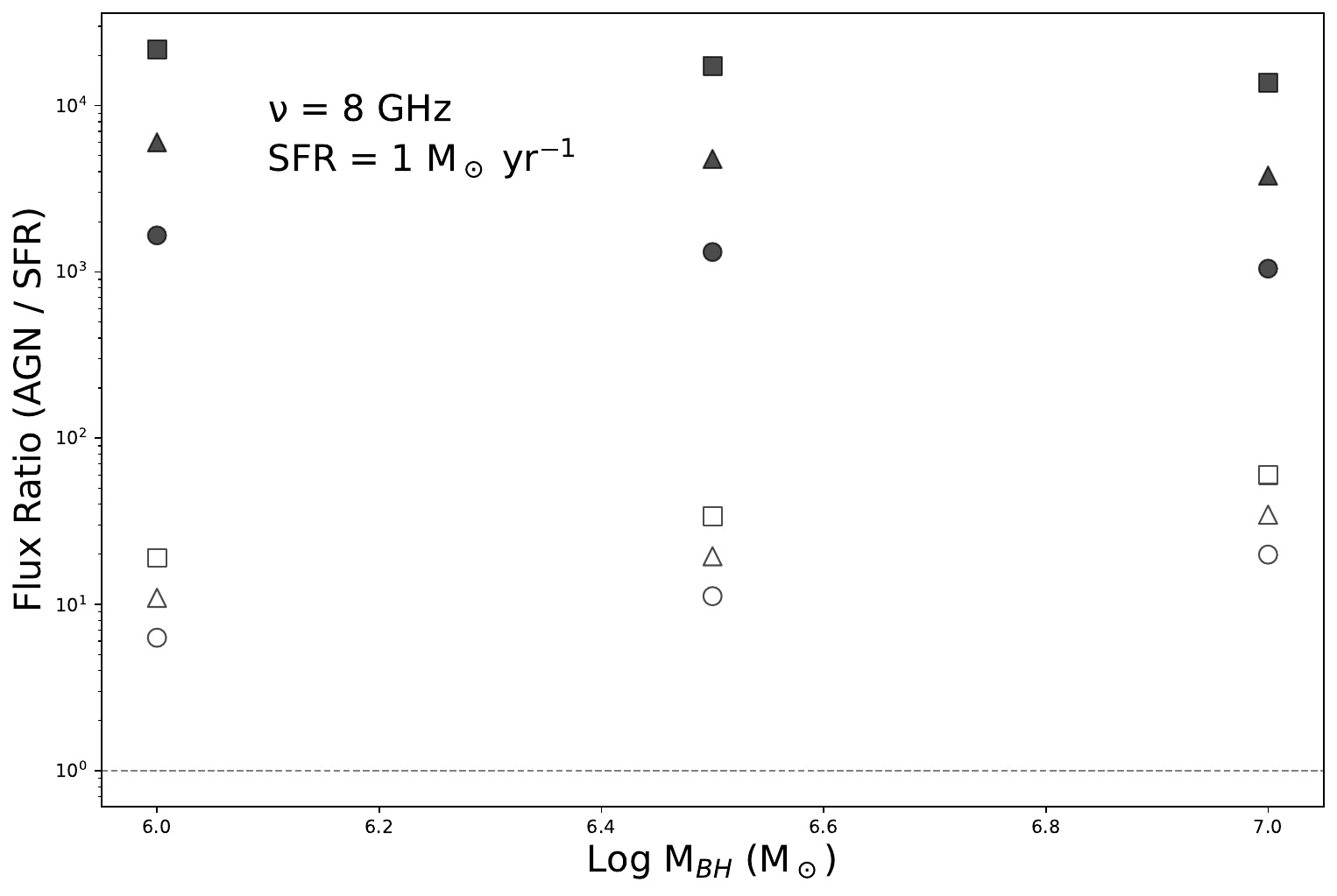} 
\includegraphics[scale=0.35]{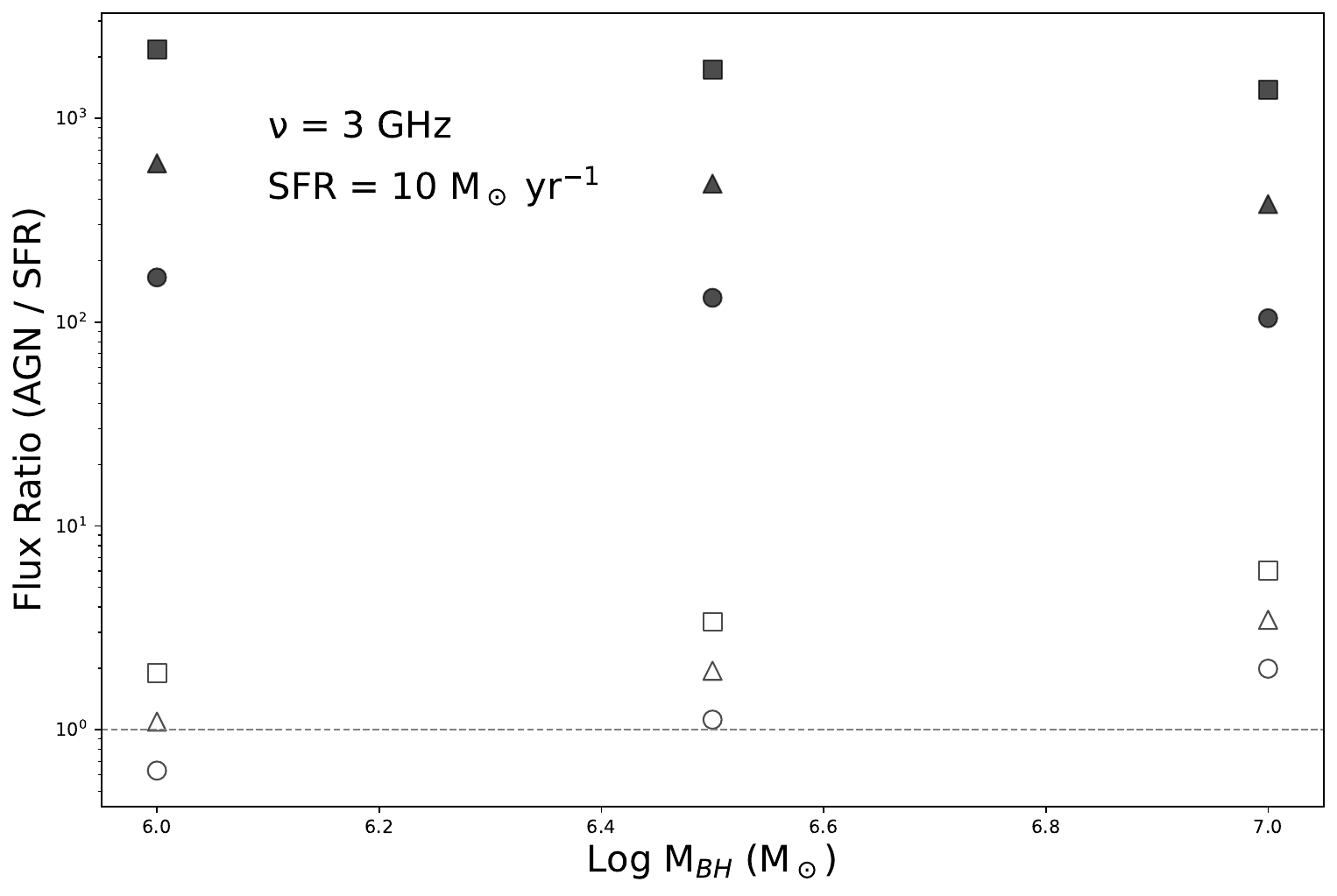} 
\includegraphics[scale=0.35]{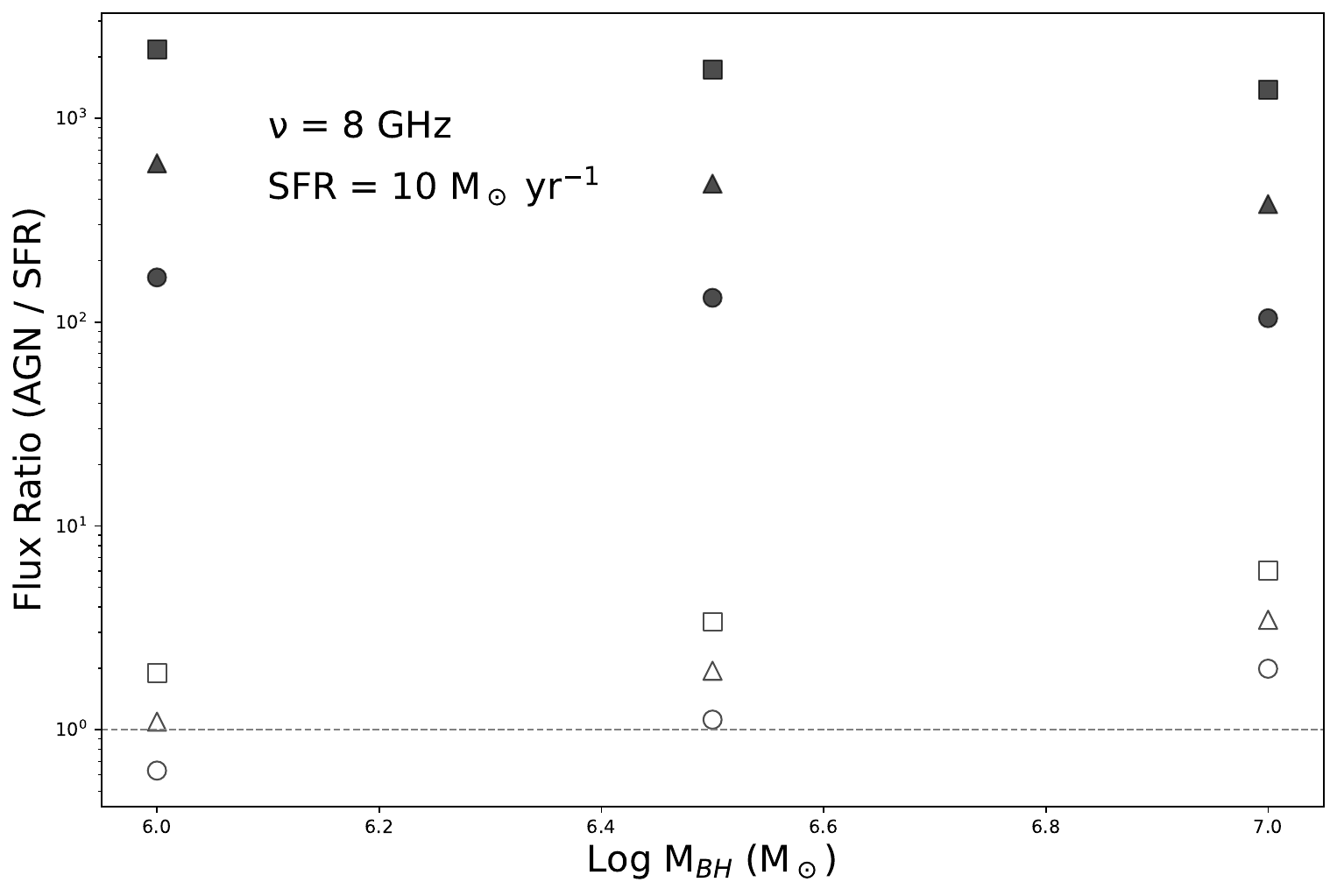} 
\includegraphics[scale=0.35]{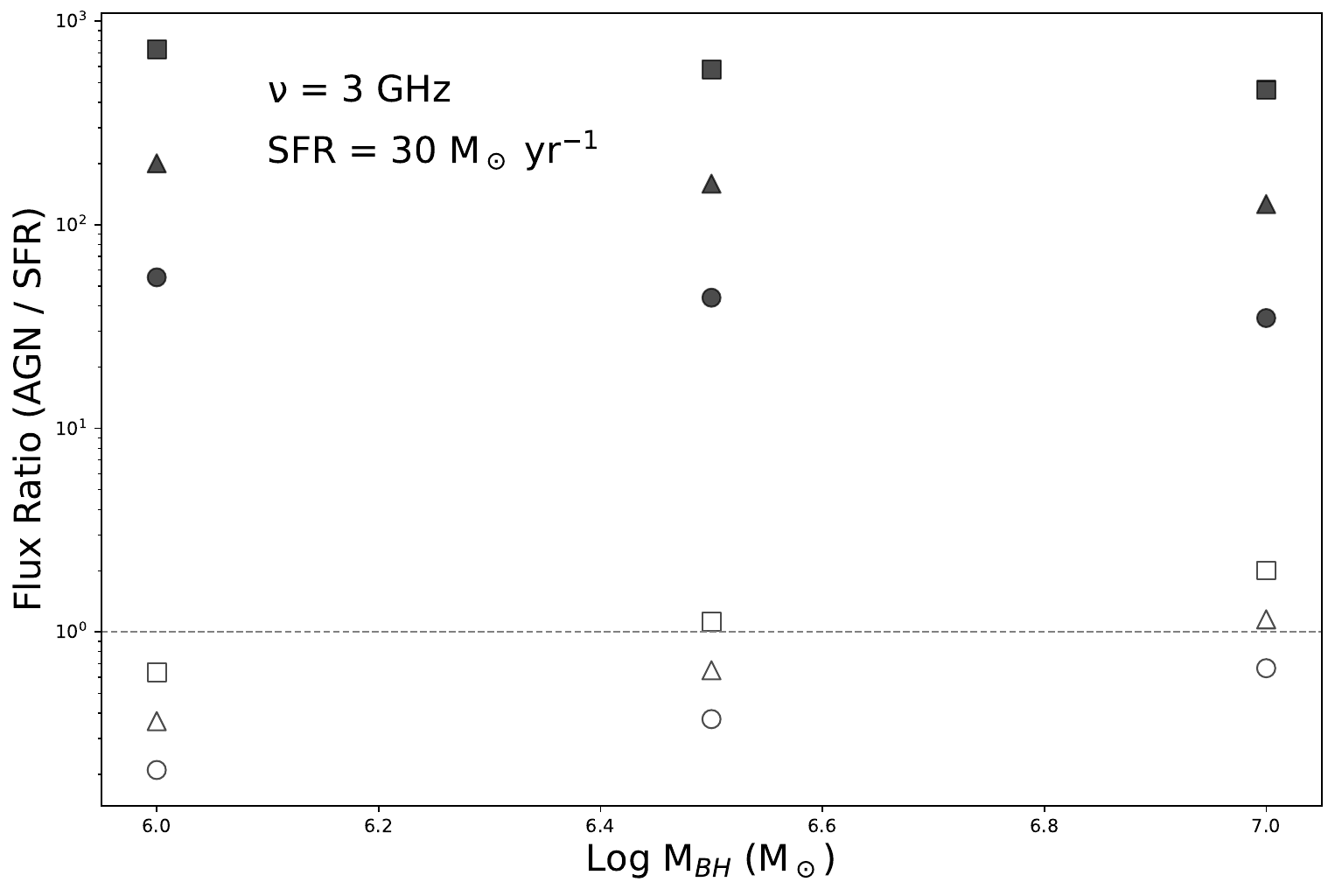} 
\includegraphics[scale=0.35]{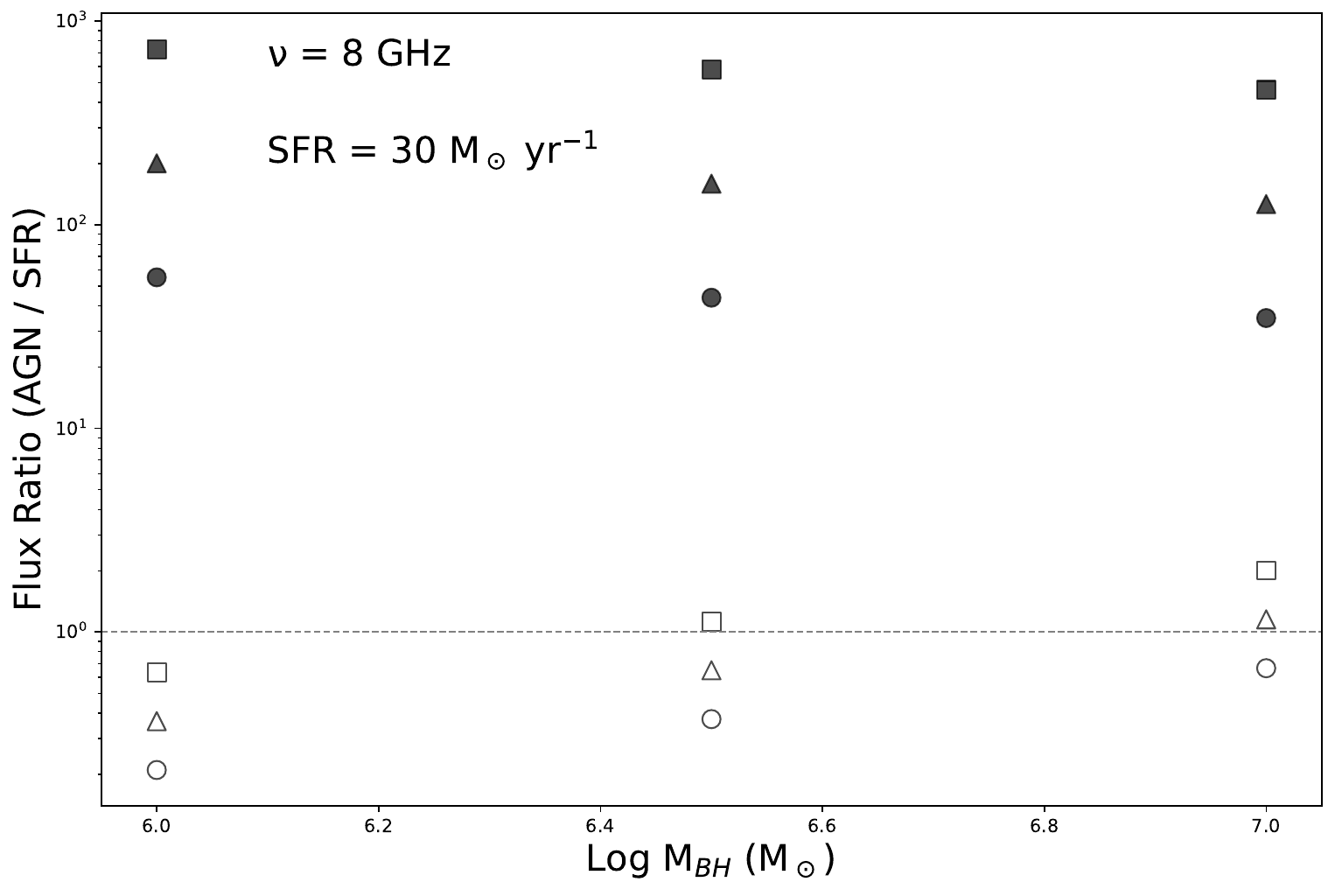} 
\end{center}
\vspace{-0.1cm}
\caption{AGN to SFR flux ratios are shown vs BH mass at $z =$ 7. The empty and solid symbols are for the radio-quiet and radio-loud sources, respectively. Circles, triangles and squares are for log($L_{\rm X})$ = 43 erg ~s$^{-1}$, log($L_{\rm X})$ = 43.5 erg ~s$^{-1}$ and log($L_{\rm X})$ = 44 erg ~s$^{-1}$, respectively. Horizontal dashed lines are for AGN/SFR flux ratio = 1.}
\label{f3}
\end{figure*}

In this scenario, we introduce radio-loudness in our analysis and report the results obtained for radio-loud sources versus radio-quiet sources
which are shown in Fig \ref{f3}. We note that flux ratios are much higher compared to the standard cases discussed above.
For SFR $=$ 1 \Ms\ yr$^{-1}$, we find that flux ratios vary from 5-2000, AGN fluxes outshine the host galaxies both for radio-loud and -quiet cases irrespective of $\rm M_{BH}$ and $L_{\rm X}$. For SFR $=$ 10 \Ms\ yr$^{-1}$, the flux ratios range from 0.5-2000, fluxes from both radio loud and quiet cases are higher than those of host galaxies for $\rm M_{BH}\geq 10^{6.5}$ \Ms, the only exception is $\rm M_{BH}= 10^{6}$ \Ms with $L_{\rm X}= 10^{43}$ $\rm ~erg~ s^{-1}$.  Even in the case of SFR $=$ 30 \Ms\ yr$^{-1}$ radio loud AGN fluxes remain dominant over the stellar fluxes from the host galaxies but radio quiet AGN fluxes become subdominant. 

\begin{figure*} 
\begin{center}
\includegraphics[scale=0.3]{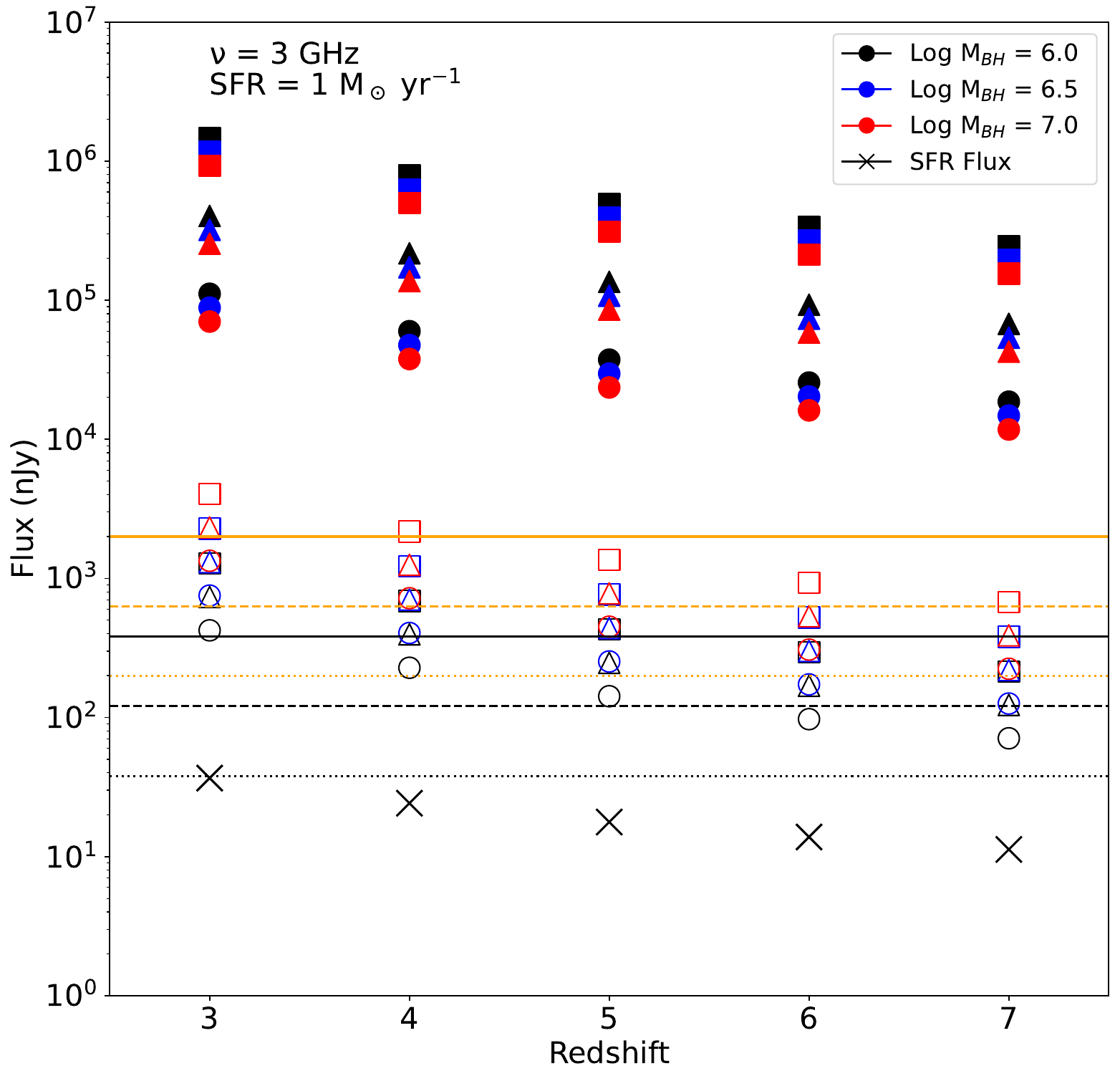} 
\includegraphics[scale=0.3]{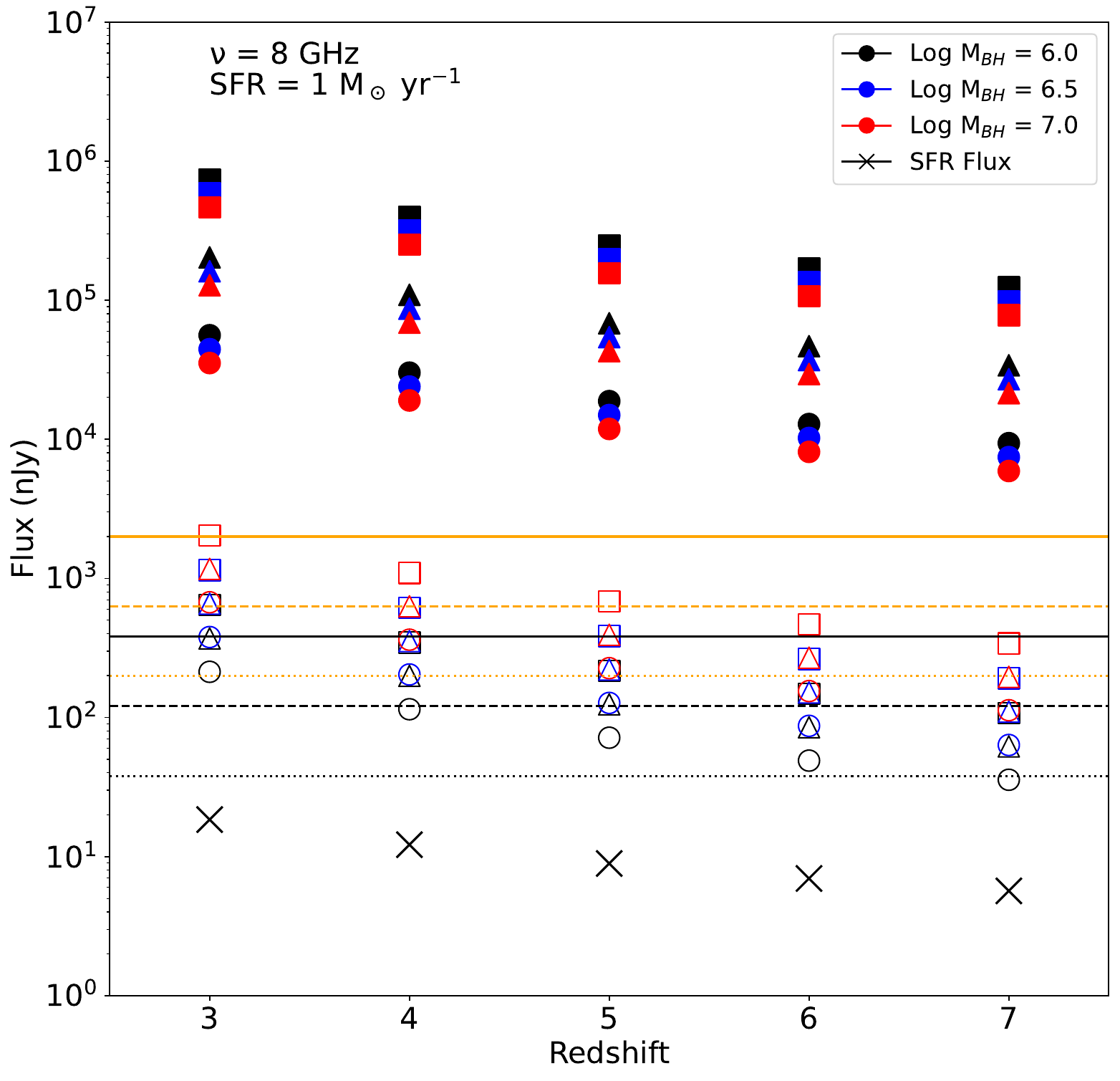} 
\includegraphics[scale=0.3]{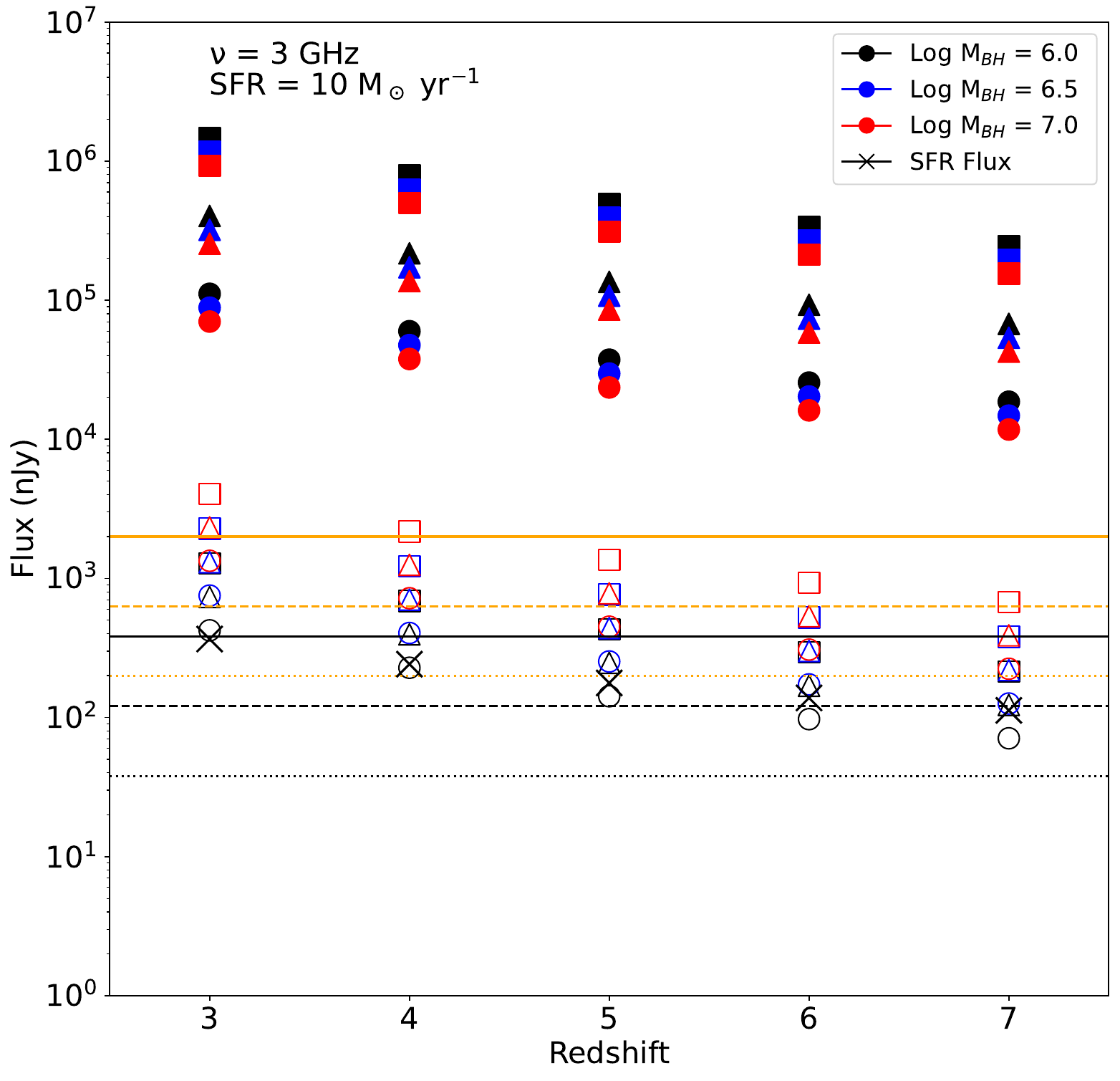} 
\includegraphics[scale=0.3]{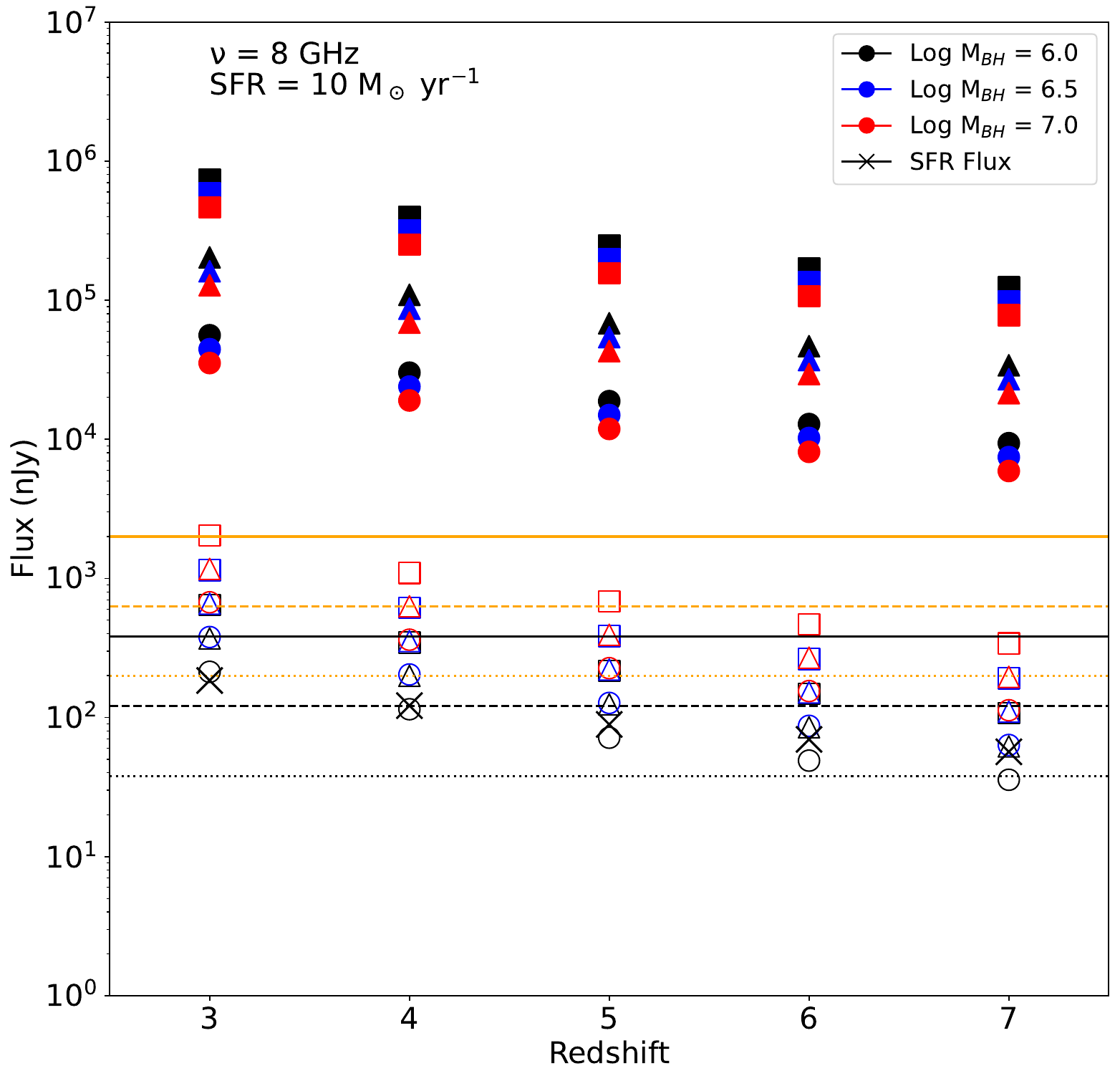} 
\includegraphics[scale=0.3]{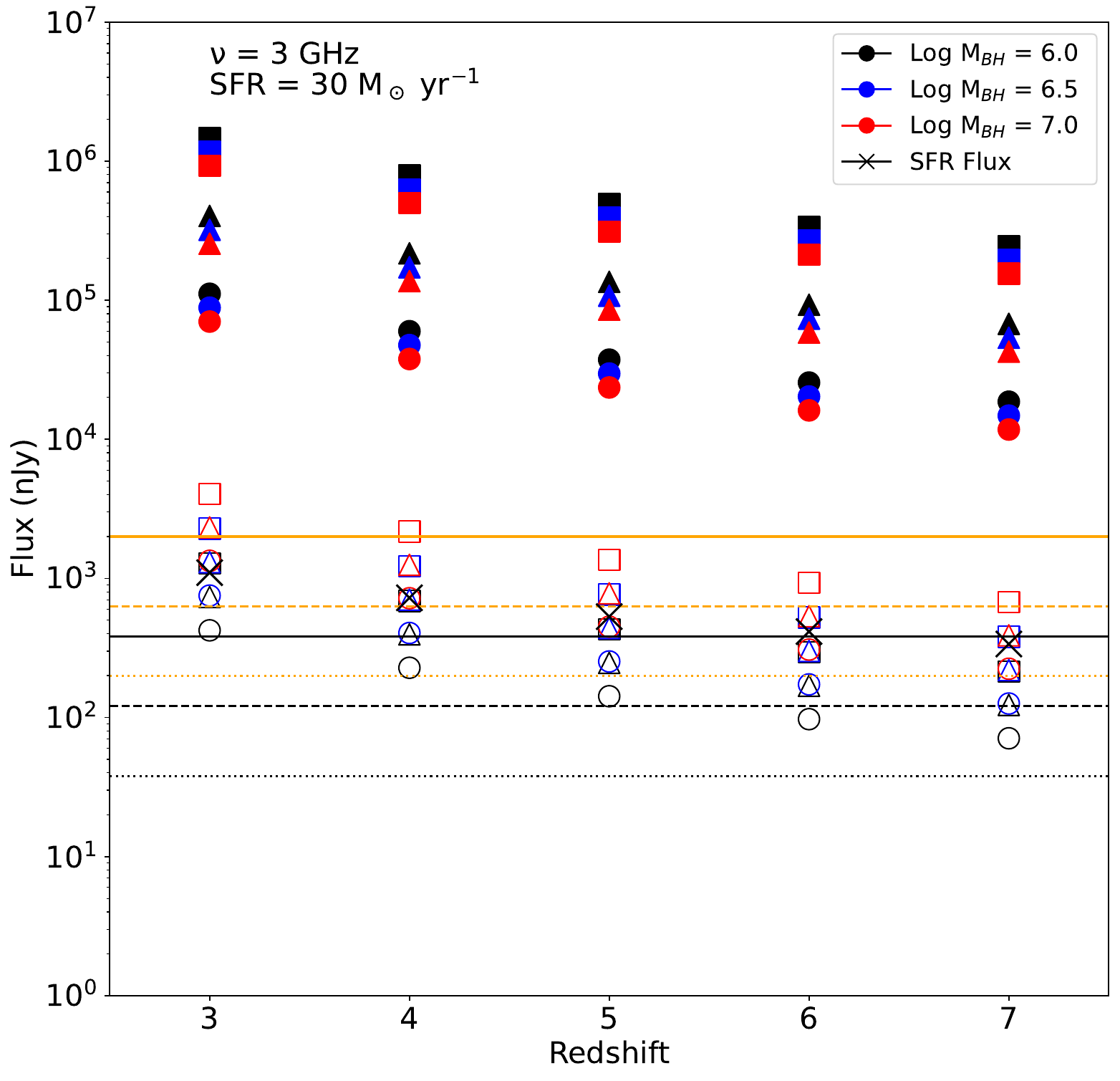} 
\includegraphics[scale=0.3]{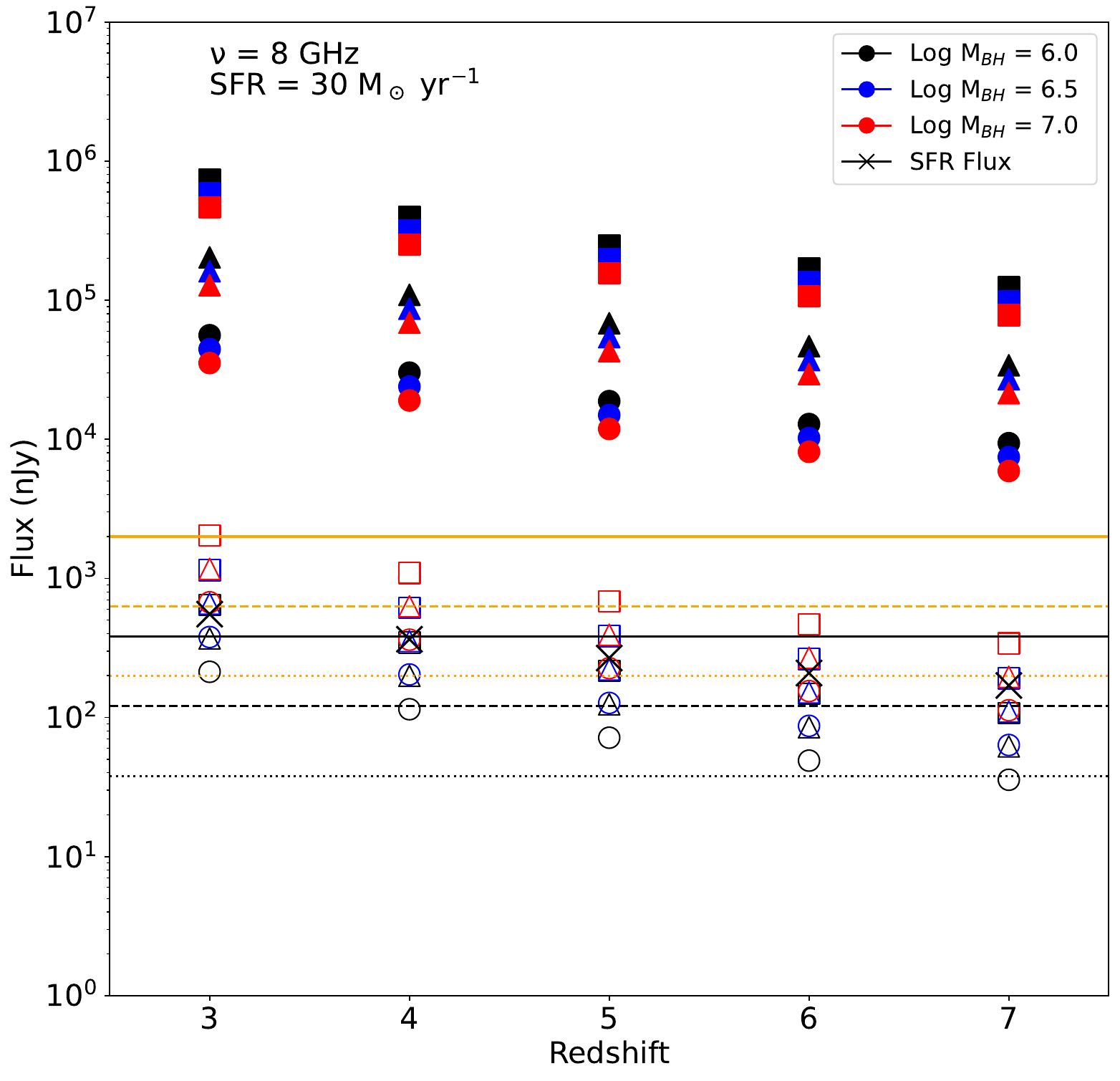} 
\end{center}
\vspace{-0.1cm}
\caption{Fluxes vs redshift. The empty symbols are for radio-quiet sources and the solid symbols for radio-loud sources. Colors indicate BH mass and black crosses show SFR fluxes.  Circles, triangles and squares are for log($L_{\rm X})$ = 43 erg ~s$^{-1}$, log($L_{\rm X})$ = 43.5 erg ~s$^{-1}$ and log($L_{\rm X})$ = 44 erg ~s$^{-1}$, respectively. The orange and black lines are sensitivity limits for SKA and ngVLA, respectively, for integration times of 1 hr (solid), 10 hr (dashed) and 100 hr (dotted).}
\label{f4}
\end{figure*}

The AGN fluxes and stellar fluxes from the host galaxies for the radio-loudness scenario are shown at $3 < z < 7$ in Fig. \ref{f4}. The radio-loud AGN fluxes range from 20,000 - 2,000,000 nJy and the radio-quiet AGN fluxes vary from 90-4000 nJy. In both cases, they only decrease by a factor of a few at $z =$ 3 - 7. The radio-quiet fluxes are about two orders of magnitude lower than the radio-loud fluxes.
The AGN radio-loud fluxes dominate the host galaxies irrespective of SFRs.
The stellar fluxes from host galaxies are subdominant than the radio-quiet AGN fluxes for SFR $=$ 1 $\Ms$  yr$^{-1}$ which are of the order of 15-40 nJy. For SFR $=$ 10 $\Ms$  yr$^{-1}$, the stellar fluxes are a few hundred nJy and become only comparable to the radio-quiet AGN fluxes only for $\rm M_{BH}= 10^{6}$ \Ms while for all of the remaining cases they are lower. In the case of SFR $=$ 30 \Ms\ yr$^{-1}$, the radio-quiet AGN fluxes only exceed the host galaxies for $\rm M_{BH}= 10^{7}$ \Ms with $L_{\rm X} = 10^{44} \rm ~erg ~s^{-1}$.

\subsection{Observational Prospects}
We find that most of the AGN fluxes can be detected by the ngVLA with integration times of about 10 hr at $3 < z < 7$ while the SKA will require approximately 100 hr. These estimates are based on the proposed reference designs, for further details see \cite{W23}. The SKA limits for 3 and 8 GHz are listed in Table 3 of \cite{B19} and the ngVLA 5-sigma values are taken from \cite{pr18}. Even 1 hr integration with the ngVLA will be able to detect AGN with $M_{\rm BH} \geq 10^{6.5}$~\Ms\ at $z < 5$. In general, for the same flux the SKA requires 10 times longer integration times than the ngVLA. Most importantly, the detection of any signal above $\sim $ 2000 nJy at $z =$ 3 - 4 and above $\sim$ 500 nJy at $z > 5$ would be a smoking gun for the presence of an AGN in LRDs if SFRs in the host galaxies are 10 $\Ms$ yr$^{-1}$. \cite{Mazo24} performed a stacking analysis of 22 JWST-selected broad line AGN and report radio non-detection down to 0.2 $\mu$Jy/beam. Similar results are found by \cite{Ak24} and \cite{G25} from the COSMOS survey. \cite{Peg25} stacked a sample of more than 900 LRDs from VLASS and FIRST surveys and obtained upper limits of $\sim$ 11 $\mu$ Jy/beam and $\sim $18 $\mu$ Jy/beam, respectively.  These observations \citep{Mazo24,Ak24,Peg25,G25} rule out the possibility that LRDs are radio-loud AGN otherwise would have been already detected in the current VLASS, FIRST and COSMOS surveys.
 The only radio detection of an LRD comes from \cite{G25}, who find a radio flux of $\sim$ 0.15 $\mu$ Jy/beam and a radio loudness of 0.5. They also found that current radio observations do not have sufficient depth to claim that JWST discovered high z AGN candidates are radio weak. This further supports our finding that the radio-quiet AGN scenario cannot be ruled out by the current observations and they would require deeper observations down to 10 nJy.

\section{Discussion and Conclusion}

We estimated radio fluxes from AGN candidates in LRDs and stellar fluxes from their host galaxies for a range of BH mass, $L_{\rm X}$, redshifts and SFRs. Also, we considered two different scenarios, the standard case and the radio-loudness case.
For the standard scenario, we found that the maximum AGN radio flux densities range from 1 - 3000 nJy and are about 100, 10 and 2 times higher than the radio emission from stars in their host galaxies for $\rm SFR = 1, 10$ and 30 \Ms\ ~yr$^{-1}$, respectively. For low SFRs of $\rm 1$~\Ms~yr$^{-1}$, AGN radio fluxes dominate stellar fluxes from their host galaxies but for higher SFRs $\geq 10$~\Ms~yr$^{-1}$ AGN fluxes only exceed those of the host galaxies for $M_{\rm BH} = 10^{7}$~\Ms. Overall, radio AGN fluxes decrease by a factor of a few from $z = 3$ to $z = 7$. For the radio-loudness scenario, radio-loud AGN fluxes are about two orders of magnitude higher compared to the corresponding radio-quiet cases. In general radio-quiet fluxes are comparable to the standard scenario. Our findings suggest that if LRDs were hosts of radio-loud AGN then they would be already detected with current observations in the VLA-COSMOS or FIRST survey.  We find that LRDs are most likely radio quiet AGN and their radio properties can be probed with the upcoming radio observatories such as ngVLA and SKA. If SFRs in the LRDs host galaxies are $\rm < 30 ~\Ms~yr^{-1}$, then the detection of radio fluxes above $\sim$ 2000 nJy at $z =$ 3 - 4 and above $\sim$ 500 nJy at $z >5 $ would be a clear indication of the presence of a BH in an LRD. Otherwise AGN radio fluxes can not be distinguished from their host galaxies. 

We have employed here the empirical relation given in \cite{D17} to estimate the radio emission from host galaxies but our estimates of the AGN to SFR flux ratios may depend on the choice of radio-SFR relation. Observations of radio sources in the local Universe suggest a median value of $\alpha$ = 0.7 \citep{ccb02} while radio emission from high-$z$ quasars indicate a median value of $\alpha$ = 0.3 \citep{glou21}. These values evince an intrinsic scatter in $\alpha$ for individual sources. Although we show here radio fluxes for $\alpha$ = 0.3, in our previous studies we found that $\alpha =$ 0.7  caused fluxes to decrease by a factor of 2 at $\nu >$ 5 GHz but increase by the same factor at low frequencies \citep[0.1 GHz;][]{W23,L24,L24b}. Our main conclusions are thus true for both values of $\alpha$. 

Some studies suggest that LRDs could be AGN accreting above the Eddington limit and are highly obscured because of large columns of gas and dust \citep{Pac24,Ina24}. Even in this case our estimates for radio emission would remain valid because radio is least affected by such effects. Moreover, in the case of super-Eddington accretion one would expect radio jets and lobes, which are expected to dissipate energy via scattering with the cosmic microwave background at high redshifts rather than emitting synchrotron radiation \citep{gh14,fg14,Gloud24}. However, if such jets are formed at lower redshifts (z$<$5) then they would be visible.

\section{acknowledgements}
  MAL thanks the UAEU for funding via SURE Plus grant No. 12S182. MM acknowledges support from the Spanish Ministry of Science and Innovation through the project PID2021-124243NB-C22. This work was partially supported by the program Unidad de Excelencia María de Maeztu CEX2020-001058-M. We thank the referee for constructive feedback.


\vspace{-50 cm}

\end{document}